\documentclass[12pt]{article}
\usepackage{amssymb}
\usepackage{amsmath,amsfonts}
\usepackage{hyperref}
\usepackage{cancel}
\usepackage[normalem]{ulem}
\usepackage{color}

\numberwithin{equation}{section}

\definecolor{blue-violet}{rgb}{0.54, 0.17, 0.89}
\definecolor{PineGreen}{cmyk}{0.92, 0, 0.59, 0.25}
\definecolor{OliveGreen}{cmyk}{0.64, 0, 0.95, 0.40}
\definecolor{RawSienna}{cmyk}{0, 0.72, 1, 0.45}
\definecolor{Gray}{cmyk}{0, 0, 0, 0.50}
\definecolor{MidnightBlue}{cmyk}{0.98, 0.13, 0, 0.43}
\definecolor{Orange}{cmyk}{0, 0.61, 0.87, 0}
\definecolor{LimeGreen}{cmyk}{0.50, 0, 1, 0}
\definecolor{Green}{cmyk}{1, 0, 1, 0}

\newcommand{\mb}{\textcolor{MidnightBlue}}

\begin{document}

\title{Asymptotic Black Holes and Conformal Mass in AdS Quadratic Curvature Gravity}

\author{Olivera Miskovic$^{1,}$\thanks{olivera.miskovic@pucv.cl} , Rodrigo Olea$^{2,}$\thanks{rodrigo.olea@unab.cl}\,,\\ Eleftherios Papantonopoulos$^{3,}$\thanks{lpapa@central.ntua.gr}\;,  and Yoel Parra-Cisterna$^{1,}$\thanks{yoel-parra@live.cl} \bigskip\\
{\small $^1$ Instituto de F\'\i sica, Pontificia Universidad Cat\'olica de Valpara\'\i so,}\\
{\small Casilla 4059, Valpara\'\i so, Chile}\\
{\small $^2$  Departamento de Ciencias Físicas, Universidad Andres Bello,}\\
{\small Sazié 2212, Piso 7, Santiago, Chile }\\
{\small $^3$ Physics Department, National Technical University of Athens,}\\
{\small 15780 Zografou Campus, Athens, Greece}\\
}

\maketitle

\begin{abstract}
We  explore the consistent truncation of conserved charges in Quadratic Curvature Gravity (QCG) with anti-de Sitter asymptotics to the linear order in the Weyl tensor.  The QCG action is given by the most general curvature-squared corrections to Einstein gravity, and it is suitably rendered finite by the addition of extrinsic counterterms (Kounterterms).
The conserved charges derived from this action are, as a consequence, nonlinear in the spacetime Riemann tensor. A detailed analysis of the falloff of generic static solutions leads to a charge proportional to the electric part of the Weyl tensor, without loss of information on the energy of the system. The procedure followed provides firmer ground to the extension of the notion of Conformal Mass to higher-curvature gravity, as it appears as associated to a renormalized action. We observe that criticality condition in QCG poses an obstruction to the charge linearization, in contrast to previous results in Lovelock gravity, where degeneracy condition plays a key role.
\end{abstract}
\newpage
\tableofcontents

\newpage

\section{Introduction}

Over the years all theoretical physical theories have been tested over their observational and experimental validity. In the case of gravity theories, their validity has to be tested against the astrophysical observations.  Differences between general relativity (GR) and alternative theories
described by modified gravity theories are expected to occur for strong gravitational fields, such as the ones created by different compact objects like neutron stars, strange stars and black holes. The $ f(R)$ and Lanczos-Lovelock gravity theories in higher dimensions have received more attention as they are the simplest generalization of GR.

To avoid restrictions from the early times cosmological observations on the gravitational Lagrangian as a linear function of $R$, variable-modified theories of gravity were investigated in the form of quadratic Lagrangians that contain some of the four possible second-order curvature invariants. One of the first such models was the Starobinsky model $f(R)=R+\alpha R^2$ \cite{Starobinsky:1980te} which was introduced to explain the inflation behavior of the early universe. In this model neutron stars were studied  \cite{Staykov:2015cfa}, in which strong gravity effects are non-negligible. It was found that, in these theories, the neutron stars can differ significantly from their GR counterpart which makes them a very good candidate to test $f(R)$ theories on astrophysical scales \cite{Doneva:2015hsa}. Also, $f(R)$ models can give important corrections in the late universe and they can lead to self-accelerating vacuum solutions, providing a purely gravitational alternative to dark energy \cite{Carroll:2003st,Carroll:2003wy,Capozziello:2002rd,Capozziello:2003gx}.

In $f(R)$ gravity theories there are black hole solutions similar to the known black hole solutions of GR or they differ considerably from their GR counterparts. Just to mention some, static spherically symmetric solutions in $f(R)$ gravity were studied in \cite{Multamaki:2006zb,delaCruzDombriz:2009et,Hendi:2011eg}, and exact spherically symmetric solutions were discussed in  \cite{Sebastiani:2010kv} different from the Schwarzschild-(A)dS solutions.

On the other hand, gravity theories that contain higher powers of Riemann curvature and still possess second-order field equations in the metric are known as Lanczos-Lovelock (LL) gravity \cite{Lanczos:1938sf,Lovelock:1971yv,Lovelock:1972vz}. In four dimensions, GR is the unique LL gravity, whereas in five dimensions there is room for Einstein-Gauss-Bonnet gravity. In higher dimensions, dimensional continuations of the Euler term are added as LL curvature polynomials.
There has been extensive work on black holes in LL gravity, which appear as the generalization of the spherically symmetric Einstein-Gauss-Bonnet black hole solutions \cite{Boulware:1985wk,Wheeler:1985nh}.

The $f(R)$ gravity with the addition of quadratic polynomials in the Riemann curvature is also referred to as Stelle gravity \cite{Stelle:1976gc}, and it is the most general quadratic in curvature modification of four-dimensional GR. In general modified theories of gravity, higher-order curvature terms have been introduced in an attempt to understand the cosmological history of the early and late cosmological evolution in a consistent way to recent observations.  On more theoretical grounds,    higher-order corrections to the Einstein-Hilbert term lead to a renormalizable and thus quantizable gravitational theory \cite{Stelle:1976gc}. Also, it was shown  in \cite{Edelstein:2021jyu} that QCG, unlike various other higher curvature theories such as Einstein-Gauss-Bonnet, is free from causality issues. Therefore, modified theories of gravity with higher-order corrections provide a deeper understanding of GR.

High derivatives of various fields appear in models, which are consistent truncations of string/M theories reduced to four spacetime dimensions. To make these models tractable \cite{Gauntlett:2009bh} constraints should be imposed on the reduction process. A typical example is the gauge/gravity duality \cite{Maldacena:1997re}, which is a powerful method of studying strongly coupled phenomena using dual weakly coupled gravitational systems (for a review, see \cite{Hartnoll:2009sz}). This duality can be considered one of the most successful applications of string theory. Higher-derivative coupling between various fields was considered in \cite{Kuang:2013oqa} in a gauge/gravity holographic model.

In holographic models described by the gauge/gravity duality, high-order curvature correction terms appear, which generate fourth-order field equations enriching the asymptotic structure of spacetime allowing the appearance of new holographic sources at the conformal boundary, in addition to new parameters to build a holographic theory modifying in this way the dynamics of the strongly coupled dual theory. In holographic hydrodynamic models, the addition of $R^{2}$ terms changes the ratio of shear viscosity over entropy density \cite{Kats:2007mq}, violating the universal bound $1/4\pi $ proposed in \cite{Kovtun:2003wp}. In \cite{Gregory:2009fj,Pan:2009xa} high-curvature terms in the form for the Gauss-Bonnet term were considered in the framework of AdS/CFT correspondence and it was found that higher curvature corrections make condensation harder and the presence of these terms violate a universal relation between the critical temperature of the superconductor and its energy gap.

In \cite{Deser:2002rt,Deser:2002jk}, a universal definition of energy was provided, and it was  evaluated  in
appropriate asymptotic geometries, for theories quadratic (or higher) in curvatures, with
or without Einstein and cosmological components. A formula for gravitational energy in covariant form was proposed in \cite{Giribet:2018hck, Giribet:2020aks} in an arbitrary theory of gravity including quadratic curvature terms in even dimensions via the Noether-Wald method \cite{Iyer:1994ys}. The basic idea is to add a topological term to the action such that it renormalizes its variation. From a finite surface term in the variation of the total action, finite asymptotic charges are derived. This procedure had already been consistently applied in second-derivative gravity theories, such as Einstein-AdS \cite{Aros:1999kt} and Einstein-Gauss-Bonnet-AdS gravity \cite{Kofinas:2006hr}.

If a charged particle is moving in an orbit in the presence of an electromagnetic field it will feel the electric or magnetic field depending on its charge. Motivated by electromagnetic field description, the Weyl tensor which represents a pure gravitational field and it is a measure of the curvature of spacetime in GR,
is decomposed into electric and magnetic parts. The electric part of the Weyl tensor contains all information about the tidal forces due to gravity, while the magnetic part contains all other information about the Weyl curvature. The
computations in GR are known to be complicated as the tensors involved are up to rank four and obtained using second-order partial derivatives of the metric tensor and their combinations.

In Einstein's gravity, the Weyl tensor is invariant under conformal transformations.
The conformal mapping between the physical metric $g_{\mu\nu}$ (a given solution of the Einstein equations) and a conformal one, $\tilde{g}_{\mu\nu}=\Omega^2 g_{\mu\nu}$, serves the purpose of defining a regular boundary,
as long as the conformal factor $\Omega$ vanishes on the boundary and its derivative is finite. In doing so, the conformal mass is encoded in the electric part of the Weyl tensor for $\tilde{g}_{\mu\nu}$.

The conformal mass is one of the conserved charges which depend on the bulk geometry and there are various ways one can define conserved charges
by studying asymptotic symmetries.   Among them, the Ashtekar-Magnon-Das \cite{Ashtekar:1984zz,Ashtekar:1999jx} (AMD) method applies Penrose's conformal transformation to determine
conserved charges in the asymptotically AdS spaces. A defining feature of this method
is that all information about the conserved charge is contained in the electric part of the Weyl tensor,
\begin{equation}
\label{eq:elWeyl}
\mathcal{E}^i_j=\frac{1}{d-2}\,W^{i\mu}_{j\nu}\,n_\mu n^\nu\,,
\end{equation}
where $n_\mu$ is the normal vector to the boundary, whose local coordinates are $x^i$.
Suitable rescaling of tensorial quantities makes the conserved charge formula expressible in terms of the Weyl tensor defined in (\ref{eq:elWeyl}).

In this work, we study the most general Quadratic Curvature Gravity (QCG) in $\left.d+1\right.$ dimensions, consisting of the Ricci-squared,  the Ricci scalar-squared and the Gauss-Bonnet (GB) terms. We first give a detailed description of QCG and we calculate the critically condition under which the physical propagating modes exist. Then we calculate the most general form of an asymptotic black hole in $(d+1)$-dimensional QCG, away from the critical point. To calculate the conformal mass, we first discuss the asymptotic form of the Weyl tensor and, to avoid possible infrared divergences, we give a detailed discussion on the Kounterterm charges in asymptotic AdS QCG. Then we give the explicit form of the conformal mass in odd and even dimensions.

The work is organized as follows. In Section \ref{QCD}, we describe the QCG theory, we derive the equations of motion and we derive the propagating modes and the criticality condition. In Section \ref{bhs}, we derive the asymptotic black hole solutions and the asymptotic form of the Weyl tensor. In Section \ref{kount}, we discuss the Kounterterm charges in the asymptotic AdS QCG. In Section \ref{conf}, we calculate the conformal mass in odd and even dimensions. In Section \ref{conc} are our conclusions and, finally, we summarize our conventions in Appendix \ref{Conventions}, and derive an asymptotic black hole solution in Appendix  \ref{ASTAdS}.


\section{Quadratic Curvature Gravity}
\label{QCD}

In this Section we will discuss the most general gravitational theory with quadratic corrections in the curvature. The action in $d+1 \geq 4$ dimensions is given by
\begin{equation}\label{QCGaction}
I_{\mathrm{QCG}}=\int\limits_{M}d^{d+1}x\sqrt{-g}\left[ \frac{1}{\kappa }%
\,(R-2\Lambda _{0})+\alpha R_{\mu \nu }R^{\mu \nu }+\beta R^{2}+\gamma \,GB\right] .
\end{equation}
It describes a modification of the standard GR in the presence of a cosmological term $\Lambda_0$, thought of as a short-range interaction. The presence of quadratic curvature contributions
consist of the Ricci-squared,  the Ricci scalar-squared   and the GB terms. The GB term is defined by
$R_{\mu\nu\alpha\beta}R^{\mu\nu\alpha\beta}-4R_{\mu\nu}R^{\mu\nu}+R^2$.

The equations of motion for the theory are obtained by performing functional variations of the action \eqref{QCGaction} with respect to the metric field, $g_{\mu \nu}(x)$, which produces
\begin{equation} \label{eomQCG}
E_{\mu \nu }=\frac{1}{\kappa }\,G_{\mu \nu }+\gamma H_{\mu \nu }+P_{\mu \nu}=0\,,
\end{equation}
where the first term is the Einstein tensor
\begin{eqnarray}\label{EinsteinTensor}
G_{\mu \nu } &=&R_{\mu \nu }-\frac{1}{2}\,g_{\mu \nu }\,R+\Lambda
_{0}\,g_{\mu \nu }\,.
\end{eqnarray}
The Gauss-Bonnet part of the action gives rise to the Lanczos tensor, $H_{\mu \nu }$,
\begin{eqnarray}\label{LanczosTensor}
H_{\mu \nu } &=&-\frac{1}{2}\,g_{\mu \nu }\left( R^{2}-4R^{\alpha \beta }R_{\alpha \beta
}+R^{\alpha \beta \lambda \sigma }R_{\alpha \beta \lambda \sigma }\right)
\notag \\
&&+2\left( RR_{\mu \nu }-2R_{\mu \lambda }R_{\ \nu }^{\lambda }-2R_{\mu
\alpha \nu \beta }R^{\alpha \beta }+R_{\mu \lambda \alpha \beta }R_{\nu }^{\
\lambda \alpha \beta }\right) \,.
\end{eqnarray}
Finally, the symmetric tensor $P_{\mu \nu }$ contains higher-derivative contributions to the field equations, which come from the  the Ricci-squared and the Ricci scalar-squared terms in the action \eqref{QCGaction}, that is,
\begin{eqnarray}\label{PTensor}
P_{\mu \nu } &=&2\beta R\left( R_{\mu \nu }-\frac{1}{4}\,g_{\mu \nu
}R\right) +\left( \alpha +2\beta \right) \left( g_{\mu \nu }\square -\nabla
_{\mu }\nabla _{\nu }\right) R  \notag \\
&&+\alpha \square G_{\mu \nu }+2\alpha \left( R_{\mu \sigma \nu \lambda }-
\frac{1}{4}\,g_{\mu \nu }R_{\sigma \lambda }\right) R^{\sigma \lambda }\,.
\end{eqnarray}
It is clear that, when $P_{\mu \nu } \neq 0$, the equations of motion of QCG are of fourth order in derivatives.


\subsection{Effective cosmological constant and degeneracy condition}
\label{Effective cosmological constant}

In the action \eqref{QCGaction}, the parameters $\alpha$, $\beta$ and $\gamma$ are introducing length scales in the theory
and therefore these scales should redefine the bare cosmological constant $\Lambda _{0}$ setting an effective cosmological constant
$\Lambda_{\textrm{eff}}$. On the other hand,
those couplings are such that they allow the existence of maximally symmetric vacuum states present in the action, which satisfy
\begin{equation}\label{RiemannQCG}
R_{\alpha \beta }^{\mu \nu }-\frac{2\Lambda _{\mathrm{eff}}}{d(d-1)}\,\delta _{\alpha \beta }^{\mu \nu }=0\,.
\end{equation}
By plugging in the condition  \eqref{RiemannQCG} into the expressions \eqref{EinsteinTensor}, \eqref{LanczosTensor} and \eqref{PTensor}, respectively,
\begin{eqnarray}
G_{\mu \nu } &=&\left( \Lambda _{0} -\Lambda_{\mathrm{eff}}\right) \,g_{\mu\nu }\,,  \qquad  H_{\mu \nu } =-\frac{2(d-2)(d-3) }{d(d-1)}\,\Lambda _{\mathrm{eff}}^{2}\,g_{\mu \nu }\,, \nonumber\\
P_{\mu \nu } &=&-\frac{2(d-3) }{(d-1)^{2}}\,\left( \alpha +(d+1)\beta\right) \,\Lambda_{\mathrm{eff}}^{2}\, g_{\mu \nu }\,,\label{Lambda_eff}
\end{eqnarray}
it is possible to make explicit the relation between the effective cosmological constant with the bare one, by virtue of eq.\eqref{eomQCG},
\begin{equation}\label{comparacionLambda}
    \frac{\Lambda_{0}}{2\kappa\Lambda^{2}_{\mathrm{eff}}}-\frac{1}{2\kappa\Lambda_{\mathrm{eff}}}=\frac{(d-3)}{(d-1)^{2}}\,\left(\alpha+(d+1)\beta\right)+\gamma\,\frac{(d-2)(d-3)}{d(d-1)}\,.
\end{equation}
The above expression implies the existence of two branches of the theory
\begin{equation}\label{solLambda}
    \frac{1}{\Lambda_{\mathrm{eff}}^{\pm}}=\frac{1}{2\Lambda_{0}}\left[ 1\pm\sqrt{1+8\kappa \Lambda _{0}\,\frac{d-3}{d-1}\left( \frac{\alpha +(d+1)\beta }{d-1}+\gamma \,\frac{d-2}{d}\right)}\right]\,.
\end{equation}
 The Einstein's branch $\Lambda_{\mathrm{eff}}^{+}$ describes a correction continuously connected to Einstein's theory, while the stringy branch $\Lambda_{\mathrm{eff}}^{-}$  contains solutions which do not reduce to the ones of Einstein gravity in the weak field limit \cite{Boulware:1985wk}.

This equation has real roots for the  effective cosmological constant, as long as the couplings satisfy the inequality
\begin{equation}\label{desigualdad}
    1+8\kappa \Lambda _{0}\,\frac{d-3}{d-1}\left( \frac{\alpha +(d+1)\beta }{d-1}+\gamma \,\frac{d-2}{d}\right)\geq0\,.
\end{equation}
The saturation of the above bound sets a point in the parameter space $\{\kappa ,\Lambda_{0},\alpha ,\beta ,\gamma \}$  such that there is a single, degenerate vacuum state, $\Lambda_{\mathrm{eff}}^{+}=\Lambda_{\mathrm{eff}}^{-}=2\Lambda_{0}$. This relation is referred to as {\it{degeneracy}} condition for QCG, and produces a class of solutions whose metric is inequivalent respect to the ones in the Einstein branch, regarding their asymptotic behavior.

 The focus of the present study are  asymptotically AdS spaces in QCG, with the effective cosmological constant given by
\begin{equation}\label{effectiveAdS}
    \Lambda_{\mathrm{eff}}=
    -\frac{d(d-1)}{2\ell_{\mathrm{eff}}^{2}}\,,
\end{equation}
in terms of the effective AdS radius $\ell_{\mathrm{eff}}$, which is a solution of the equation
\begin{equation}\label{Eff-radius}
\frac{1}{\ell _{0}^{2}}=\frac{1}{\ell _{\mathrm{eff}}^{2}}-\frac{\kappa
d\left( d-3\right) }{\ell _{\mathrm{eff}}^{4}}\left( \frac{\alpha
+(d+1)\beta }{d-1}+\gamma \,\frac{d-2}{d}\right) \,.
\end{equation}

In order to understand better the special points in the parameter space in higher-curvature gravities, in particular in QCG, it is useful to introduce a parameter that characterizes the degeneracy of the  gravitational vacuum state, including its multiplicity.

In general, for a Lagrangian of the form $\mathcal{L}(R_{\alpha \beta }^{\mu
\nu })$, with $E_{\mu \nu }^{\alpha \beta }=\partial \mathcal{L}/\partial R_{\alpha \beta }^{\mu \nu }$, the effective AdS radius is
obtained from the equation
\begin{equation}
\left( 1-d+2\ell _{\mathrm{eff}}^{-2}\,\frac{d}{d\ell _{\mathrm{eff}}^{-2}}\right) E_{\alpha \beta }^{\alpha \beta }|_{\mathrm{AdS}}=0\,.
\label{effective AdS}
\end{equation}
When the Lagrangian contains the Einstein's term $R=\frac{1}{2}\,\delta _{\mu\nu }^{\alpha \beta }R_{\alpha \beta }^{\mu \nu }$, such that $\mathcal{L}=\frac{\kappa }{2}\,\delta _{\mu \nu }^{\alpha \beta }R_{\alpha \beta }^{\mu
\nu }+\cdots $, it is convenient to define  the polynomial proportional to eq.\eqref{effective AdS}, and normalized as
\begin{equation}
\Theta _{d}\equiv -\frac{2\kappa }{(d+1)d(d-1)}\left( 1-d+2y\,\frac{d}{dy}\right) E_{\alpha \beta }^{\alpha \beta }|_{\mathrm{AdS}}\,,
\end{equation}
which has the same degree in $y=\ell _{\mathrm{eff}}^{-2}$ as a degree of $\mathcal{L}$ is in the Riemann curvature. Then the solution $y_{k}$ corresponds to a degenerate vacuum that has the multiplicity $k$ if the first non-vanishing derivative of $\Theta _{d}$ in $k$ is $\Theta _{d}^{(k)}\equiv \left. \frac{d^{k}\Theta _{d}}{dy^{k}}%
\right\vert _{y_{k}}$. In particular, in Lovelock gravity, $\Theta
_{d}^{(1)}=0$ is an obstruction to linearize a theory and also to define the Conformal Mass, because the coefficient $\Theta _{d}^{(1)}$ appears as an overall factor in the electric part of the Weyl tensor \cite{Jatkar:2015ffa,Arenas-Henriquez:2019rph}.

For QCG theory, the double trace of the auxiliary tensor is
\begin{equation}
E_{\alpha \beta }^{\alpha \beta }|_{\mathrm{AdS}}=\left( d+1\right) d\left[
\frac{1}{2\kappa }-\frac{1}{\ell _{\mathrm{eff}}^{2}}\left( \rule{0pt}{13pt}%
\alpha d+\beta \left( d+1\right) d+\gamma \left( d-1\right) \left(
d-2\right) \right) \right] ,
\end{equation}
such that the polynomial $\Theta _{d}$ is quadratic in $\ell _{\mathrm{eff}}^{-2}$,
\begin{equation}
\Theta _{d}=-\frac{1}{\ell _{0}^{2}}+\frac{1}{\ell _{\mathrm{eff}}^{2}}-\frac{\kappa }{\ell _{\mathrm{eff}}^{4}}\frac{d\left( d-3\right) }{d-1}\left( \alpha +(d+1)\beta +\gamma \,\frac{(d-1)(d-2)}{d}\right) \,.
\label{theta}
\end{equation}
Then,  the degeneracy of the AdS vacuum is determined by the coefficient
\begin{equation}
\Theta _{d}^{(1)}
=1-\frac{d(d-3)}{d-1}\,\omega _{d}\,,   \label{degeneracy}
\end{equation}
where the shorthand
\begin{equation}
\omega _{d}=\frac{2\kappa }{\ell _{\mathrm{eff}}^{2}}\,\left( \alpha
+(d+1)\beta +\gamma \,\frac{(d-1)(d-2)}{d}\right) \,, \label{omega}
\end{equation}
is introduced as it will be of often use below.

It is straightforward to check that, when $\Theta _{d}^{(1)}=0$, the coupling constants are not independent and the effective AdS radii coincide at the value $\ell _{\mathrm{eff}}=\ell _{0}/\sqrt{2}$.


\subsection{Propagating modes and criticality condition}

In order to identify the propagating degrees of freedom of the theory, the linearized field equations for QCG are analyzed in this section. In general, quadratic-curvature corrections in the  action \eqref{QCGaction}, produce a massive tensorial mode, namely a massive graviton\footnote{For more information on the propagating modes in quadratic gravity, see reference \cite{Bueno:2016ypa}.}. The corresponding mass of this graviton will give rise to the notion of criticality in the theory \cite{Lu:2011zk, Deser:2011xc},  which will also impact in the definition of Conformal Mass in the theory.

When small perturbations of the spacetime metric, $g_{\mu \nu}$, around a maximally symmetric background, $\tilde{g}_{\mu \nu}$, are considered, that is,
 \begin{equation}\label{lineal}
     g_{\mu \nu}=\tilde{g}_{\mu \nu}+ h_{\mu \nu}+\mathcal{O}(h^{2})\,,
 \end{equation}
the linearized version of the EOM can be obtained following the procedure shown in ref.\cite{Deser:2011xc}. The result is an expansion in different orders in the metric fluctuation,
\begin{equation}
    E_{\mu \nu}=\tilde{E}_{\mu \nu}+E^{(1)}_{\mu \nu}+\mathcal{O}(h^{2})=0\,,
\end{equation}
where $\tilde{E}_{\mu \nu}=0$ is satisfied identically in the vacuum, what leaves the linear contribution in the form
\begin{eqnarray}
E_{\mu \nu }^{(1)} &=&\frac{a}{\kappa }\,G_{\mu \nu }^{(1)}+(\alpha +2\beta
)\left( \tilde{g}_{\mu \nu }\tilde{\Box}-\tilde{\nabla}_{\mu }\tilde{\nabla}%
_{\nu }-\frac{d}{\ell _{\mathrm{eff}}^{2}}\,\tilde{g}_{\mu \nu }\right)
R^{(1)}\,  \notag  \label{LinearizedEOM} \\
&&+\alpha \left( \tilde{\Box}G_{\mu \nu }^{(1)}+\frac{d-1}{\ell _{\mathrm{eff}}^{2}}\,\tilde{g}_{\mu \nu }R^{(1)}\right) =0\,.
\end{eqnarray}
Here, all differential operators with tilde are defined with respect to the background metric and the constant $a$ is given by
\begin{equation}\label{constantC}
a=1-d\omega _{d}+\frac{2\kappa }{\ell _{\mathrm{eff}}^{2}}\,\left( \rule{0pt}{12pt}\alpha -2(d-2)\gamma \right) \,.
\end{equation}
Taking the trace of the linearized equation of motion \eqref{LinearizedEOM}, and upon the gauge choice $\tilde{\nabla}^{\mu}h_{\mu \nu}=\tilde{\nabla}_{\nu}h$, a wave equation is obtained for $h=\tilde{g}^{\mu \nu}h_{\mu \nu}$
\begin{equation}\label{linerizedtrace}
\left[ \left( \rule{0pt}{12pt}(d+1)\alpha +4d\beta \right) \tilde{\Box}+%
\frac{1-d+d(d-3)\,\omega _{d}}{\kappa }\right] h=0\,.
\end{equation}
The trace of the perturbation, $h$, becomes a propagating mode,  which corresponds to a scalar field whose dynamics is governed by eq.\eqref{linerizedtrace} and which, in general,  has a nonzero mass with respect to the background. This massive scalar mode can be eliminated by imposing constraint on the parameters given by $(d+1)\alpha +4d\beta=0$, resulting in $h=0$. This condition implies that a gauge choice reduces to the transverse one, $\tilde{\nabla}^{\mu}h_{\mu \nu}=0$, such that the linearized equation \eqref{LinearizedEOM}  becomes \cite{Deser:2011xc, Gullu:2011sj}
\begin{equation}
\alpha \left( \tilde{\Box}+\frac{2}{\ell _{\mathrm{eff}}^{2}}-M^{2}\right)
\left( \tilde{\Box}+\frac{2}{\ell _{\mathrm{eff}}^{2}}\right) h_{\mu \nu
}=0\,,  \label{4waveequation}
\end{equation}
where the constant $M$ takes the value
\begin{equation}
M^{2}=\frac{2}{\ell _{\mathrm{eff}}^{2}}-\frac{a}{\alpha \kappa }\,.
\label{gravitonmass}
\end{equation}
The equation \eqref{4waveequation} describes a  massless spin-2 particle, $h^{(m)}_{\mu \nu}$, which satisfies
\begin{equation}\label{masslesswave}
    \left(\tilde{\Box} +\frac{2}{\ell_{\mathrm{eff}}^{2}}\right)  h^{(m)}_{\mu \nu}=0 \,,
\end{equation}
and a massive spin-2 propagating mode, $h^{(M)}_{\mu \nu}$,
\begin{equation}\label{massivewave}
    \left(\tilde{\Box}+\frac{2}{\ell _{\mathrm{eff}}^{2}}-M^{2}\right)h^{(M)}_{\mu \nu}=0\,.
\end{equation}
Both modes are transverse and traceless ($h^{(m)}=h^{(M)}=0$). For an AdS background,  the stability of the massive spin-2 mode requires that $M^{2}\geq0$, as shown, e.g., in \cite{Myung:2013uka}.

It is important to stress that the existence of the massive graviton depends mainly on the Ricci-squared term in the action \eqref{QCGaction}.  By combining eqs.\eqref{constantC} and \eqref{gravitonmass}, it is shown that the mass of the graviton takes the form
\begin{equation}\label{relationmassgraviton}
    M^{2}=-\frac{1}{\alpha\kappa}\,\Xi_{d}\,,
\end{equation}
where the new constant $\Xi _{d}$ is defined as
\begin{equation}
\Xi _{d}=1-\frac{2\kappa d}{\ell _{\mathrm{eff}}^{2}}\left( \alpha
+(d+1)\beta +\gamma \,\frac{(d-2)(d-3)}{d}\right) \,,
\end{equation}
which, in turn, can be rewritten in terms of  eq.\eqref{omega} as
\begin{equation}
\Xi _{d}=1-d\omega _{d}+\frac{4(d-2)}{\ell _{\mathrm{eff}}^{2}}\,\kappa
\gamma \,. \label{Xi_d}
\end{equation}
When set to zero, the criticality condition $\Xi_d=0$ imposes a constraint on the parameters of the theory,  such that the massive graviton turns massless with respect to the vacuum.
When the above condition is met, the linearized equation of motion \eqref{4waveequation} reduces to
\begin{equation}
    \left(\tilde{\Box}+\frac{2}{\ell _{\mathrm{eff}}^{2}}\right)^{2}h_{\mu \nu}=0\,,
\end{equation}
which is a sort of Klein-Gordon-squared equation. As discussed in \cite{Alishahiha:2011yb}, it corresponds to a fourth-order differential equation, which exhibits degenerate solutions and, therefore, logarithmic modes which appear naturally.

To conclude, the functions of the couplings which are relevant in QCG are the ones that defines the degeneracy condition, $\Theta^{(1)}_d$, and  the criticality condition, $\Xi _{d}$. In the next sections we will discuss their role in the definition of conserved charges.


\section{Asymptotic form of the black hole metric}
\label{bhs}

In this Section, we will discuss generic black hole solutions in QCG.
In that respect, static, Schwarzschild-like, black hole solutions in quadratic curvature gravity, with and without cosmological constant, have been a subject of study during the last few decades  (see, e.g., \cite{Stelle:1977ry,  Kehagias:2015ata, Stelle:2017bdu, Svarc:2018coe}). In the context of asymptotically AdS gravity, topological black hole solutions generalize the geometry of the transversal section from a sphere to planar or hyperbolic surfaces \cite{Birmingham:1998nr, Mann:1997iz}.

A static black hole ansatz in the local coordinates $x^{\mu} = (t, r, \varphi^m)$, is described by the metric
\begin{equation}\label{ansatzTBH}
ds^{2}=-f^{2}(r)dt^{2}+\frac{dr^{2}}{f^{2}(r)}+r^{2}\gamma _{mn}(\varphi )\,d\varphi ^{m}d\varphi ^{n}\,,\quad
\varphi ^{m}\in \Gamma ^{d-1}\,,
\end{equation}
where $\gamma_{m n}(\varphi)$ is the metric of a $(d-1)$-dimensional Riemann space $\Gamma^{d-1}$ with constant curvature $k$,
\begin{equation}\label{constanttrasnversalcurvature}
    \mathcal{R}^{mn}_{pq}(\gamma)=k\,\delta^{mn}_{pq}\,,
\end{equation}
with the parameter $k= +1$, $0$ and $-1$ that locally describes a spherical, flat and hyperbolic transversal topology, respectively. The spacetime between the horizon and infinity is foliated by a radial coordinate adapted to the topology of the transversal section and, in particular, the one of the horizon $r = r_{+}$.

In order to find the asymptotic form of a static black hole \eqref{ansatzTBH} in QCG, the first integral of the equations of motion \eqref{eomQCG} should be found. In the ansatz \eqref{ansatzTBH}, there is only one independent component of the field equations, and that is why one may analyze $E^t_t=0$,  which consists of
\begin{equation}\label{Einsteinder}
\frac{2r^{d-1}}{d-1}\,G_{t}^{t} =\left[ r^{d-2}\left( f^{2}-k\right) -\frac{r^{d}}{\ell _{0}^{2}}\right] '\,,
\end{equation}
\begin{equation}\label{Lanczosder}
    \frac{2r^{d-1}}{d-1}\,H_{t}^{t} =\left[ -(d-2)(d-3)\,r^{d-4}\left(
f^{2}-k\right) ^{2}\right] ',
\end{equation}
where the prime denotes a radial derivative. Similarly, the evaluation of the higher-order term $P_{t}^{t}$ produces
\begin{eqnarray}\label{Psi}
  \frac{2r^{d-1}}{d-1}\,P_{t}^{t}&=& \frac{2r^{d-1}}{d-1}\,2\beta R\left(
G_{t}^{t}|_{\Lambda_{0}=0}+\frac{1}{4}\,R\right) +\left( \frac{\alpha }{2}\,+2\beta \right)
\frac{2r^{d-1}}{d-1}\,\square R  \notag \\
&&+\alpha \,\frac{2r^{d-1}}{d-1}\,\square R_{t}^{t}+2\alpha \frac{2r^{d-1}}{%
d-1}\,\left( R_{t\lambda }^{t\sigma }-\frac{1}{4}\,R_{\lambda }^{\sigma
}\right) R_{\sigma }^{\lambda } \equiv [\Psi(r)]'\,,
\end{eqnarray}
where $\Psi(r)$ is a function to be determined and the Einstein tensor without cosmological constant contributes with
\begin{equation}\label{G0}
\frac{2r^{d-1}}{d-1}\,G_{t}^{t}|_{\Lambda_{0}=0}=\left[ \rule{0pt}{12pt}r^{d-2}\left( f^{2}-k\right) \right]'.
\end{equation}

Putting together the expressions \eqref{Einsteinder}, \eqref{Lanczosder} and \eqref{Psi}, the $E^{t}_{t}$ component of the equation of motion \eqref{eomQCG} can be written as a total derivative
\begin{equation}\label{derivativeE}
\frac{2r^{d-1}}{d-1}\,E_{t}^{t}=\left[ \frac{1}{\kappa }\,r^{d-2}\left(
f^{2}-k\right) -\frac{r^{d}}{\kappa \ell _{0}^{2}}-\gamma
(d-2)(d-3)\,r^{d-4}\left( f^{2}-k\right) ^{2}+\Psi \right]'=0\,,
\end{equation}
such that its first integral gives rise to the mass parameter, $\mu $, as an integration constant
\begin{equation}\label{first integral}
\frac{1}{\kappa }\,\left( f^{2}-k\right) -\frac{r^{2}}{\kappa \ell _{0}^{2}}-\gamma (d-2)(d-3)\,\frac{\left( f^{2}-k\right) ^{2}}{r^{2}}+\frac{\Psi }{r^{d-2}}=-\frac{\mu }{r^{d-2}}\,.
\end{equation}
In a gravity theory with arbitrary quadratic-curvature couplings, it is difficult to find the exact form of the function $\Psi(r)$. Therefore, only an asymptotic analysis of the equations of motion can be performed.

\subsection{Degeneracy {\it{vs}} Criticality}

In order to illustrate the effect of curvature-squared terms in the action over the behavior of the metric function, one may consider
topological black holes in Einstein-Gauss-Bonnet AdS gravity (QCG with $\alpha =\beta =0$). The exact solution corresponds to the Boulware-Deser black hole whose metric function, in the Einstein's branch of theory, is given by \cite{Boulware:1985wk}
\begin{equation}\label{EGBmetricfunction}
f^{2}_{\mathrm{EGB}}=k+\frac{r^{2}}{2\kappa \gamma \left( d-2\right) \left( d-3\right) }\left[ 1-\sqrt{1-4\kappa \gamma \left( d-2\right) \left( d-3\right) \left(
\frac{1}{\ell _{0}^{2}}-\frac{\kappa\mu }{r^{d}}\right) }\right] \,.
\end{equation}
The above function, when expanded for $r\to\infty$, adopts the form
\begin{equation} \label{EGBexpansion}
f^{2}_{\mathrm{EGB}}=k+\frac{r^{2}}{\ell _{\mathrm{eff}}^{2}}-\frac{\mu\kappa }{\Xi _{d}\,r^{d-2}}+\frac{\mu ^{2}\kappa^{3} \gamma (d-2)(d-3)}{\Xi _{d}^{3}\,r^{2d-2}}+\mathcal{O}\left( \frac{1}{r^{3d-2}}\right) \,,
\end{equation}
where the falloff of the mass term is dictated by the constant \begin{equation}
  \Xi _{d}=1-\frac{2\kappa
\gamma (d-3)(d-2)}{\ell _{\mathrm{eff}}^{2}} \,,
\end{equation}
which corresponds to the criticality condition defined in eq.\eqref{Xi_d}, but for $\alpha=\beta=0.$ Remarkably enough, in EGB gravity, the notion of criticality  coincides with the concept of degeneracy, i.e., $\Xi _{d}=\Theta_{d}^{(1)} $. Therefore, either condition poses an obstruction to a Schwarzschild-like asymptotic behavior in eq.\eqref{EGBexpansion}. The role of degeneracy in EGB AdS gravity has been properly emphasized in the linearization of the conserved charges in \cite{Jatkar:2015ffa}: Conformal Mass cannot be defined in the case both effective AdS radii coalesce. A similar feature is made manifest by the general falloff of the metric function in Lovelock AdS gravity, as discussed in \cite{Arenas-Henriquez:2019rph}.

It is reasonable then to expect that there exists a class of solutions with an expansion as for asymptotic Schwarzschild-Tangherlini-AdS black holes
\begin{equation}\label{falloff}
f^{2}=k+\frac{r^{2}}{\ell_{\mathrm{eff}}^{2}}-\frac{m}{r^{d-2}}+\frac{p}{r^{2d-2}}+\mathcal{O}\left( \frac{1}{r^{2d-1}}\right) \,,
\end{equation}
where $m$ and $p$ are real constants yet to be determined. This argument excludes the existence of logarithmic terms in a relaxed asymptotically AdS behavior on critical points of the theory \cite{Alishahiha:2011yb}. 

Assuming the expansion \eqref{falloff}, the asymptotic resolution of the field equations for QCG leads to the function $\Psi(r)$, which is \footnote{The integration constant $\Psi _{0}$ may be fixed only by nonlocal considerations. Indeed, the value
\begin{equation}\label{Psi_0}
    \Psi _{0}=\frac{2d\left( \alpha +(d+1)\beta \right) m}{\ell _{\mathrm{eff}}^{2}}\,,
\end{equation}
appears as coming from a detailed discussion on (global) conserved charges in later sections.
}
\begin{equation}
\Psi=\Psi _{0}-\frac{d(d-3)\left( \rule{0pt}{12pt}\alpha +\left( d+1\right)
\beta \right) }{(d-1)\ell _{\mathrm{eff}}^{4}}\,r^{d}+\frac{2d\left( \rule{0pt}{12pt}(5d-1)\alpha +3\left( d-1\right) \beta \right) p}{\ell _{\mathrm{eff}}^{2} \,r^{d}} +\mathcal{O}\left( \frac{1}{r^{d+1}}\right) \,.
\end{equation}

Notice that the term coming from the derivatives in the curvature appears as sub-leading contributions with respect to the ones responsible for the mass. The first integral in eq.\eqref{first integral} can be expanded as
\begin{eqnarray}\label{expandintegral}
-\frac{\mu }{r^{d-2}} &=&\left[ \frac{1}{\kappa }\left( \frac{1}{\ell_{\mathrm{eff}}^{2}} -\frac{1}{\ell _{0}^{2}}\right) -\frac{\gamma (d-2)(d-3)}{\ell _{\mathrm{eff}}^{4}}-\frac{d(d-3)\left( \alpha +\left( d+1\right) \beta
\right) }{(d-1)\ell_{\mathrm{eff}}^{4}}\right] r^{2}  \notag \\
&&-\left[ 1-\frac{2\kappa d(d-3)}{\ell _{\mathrm{eff}}^{2}}\left( \frac{\ell
_{\mathrm{eff}}^{2}\Psi _{0}}{2d(d-3)m}+\frac{\gamma (d-2)}{d}\right) \right]
\frac{m}{\kappa r^{d-2}}-\frac{m^{2}\gamma (d-2)(d-3)}{r^{2d-2}} \notag \\
&&+\left[ 1-\frac{2\kappa d(d-3)}{\ell_{\mathrm{eff}}^{2}} \left( -\frac{(5d-1)\alpha +3\left( d-1\right) \beta }{d-3}+\frac{\gamma (d-2)}{d}\right)
\right] \frac{p}{\kappa r^{2d-2}}+\cdots \,. \notag
\end{eqnarray}
As expected, the first line is eliminated by recalling the definition of the effective AdS radius $\ell _{\mathrm{eff}}$, \eqref{Eff-radius}. As for the subleading order $\mathcal{O}(1/r^{d-2})$, it determines the mass parameter in terms of the integration constant,
\begin{equation}\label{massparemeter}
\mu =\left[ 1-\frac{2\kappa d}{\ell _{\mathrm{eff}}^{2}}\left( \alpha+(d+1)\beta+\frac{\gamma (d-2)(d-3)}{d}\right) \right] \frac{m}{\kappa }=\frac{\Xi_{d}}{\kappa}\,m\,,
\end{equation}
using the corresponding value in eq.\eqref{Psi_0}. Finally, the  order $\mathcal{O}(1/r^{2d-2})$ leads to
\begin{equation}
\left[ \Xi _{d}+\frac{2\kappa d\left(5d\alpha +2(2d-1) \beta \right) }{\ell _{\mathrm{eff}}^{2}}\right] \frac{p}{\kappa }=m^{2}\gamma
(d-2)(d-3)\,,
\end{equation}
such that
\begin{equation}\label{e}
p=\frac{m^{2}\kappa \gamma (d-2)(d-3)}{\Xi _{d}+\frac{2\kappa d }{\ell _{\mathrm{eff}}^{2}}\,\left( 5d\alpha +2(2d-1) \beta \right)}\,.
\end{equation}

Summing up the above results, the expansion of the metric function in a generic QCG theory is
\begin{equation}\label{metric function}
f^{2}=k+\frac{r^{2}}{\ell _{\mathrm{eff}}^{2}}-\frac{\mu \kappa }{\Xi _{d}\,r^{d-2}}+\frac{\mu ^{2}\kappa ^{3}\gamma (d-2)(d-3)}{\Xi _{d}^{2}\left( \Xi _{d}+\frac{2\kappa d}{\ell _{\mathrm{eff}}^{2}}\,\left( 5d\alpha +2\left( 2d-1\right) \beta \right) \right) r^{2d-2}}+\mathcal{O}\left( \frac{1}{r^{2d-1}}\right) \,.
\end{equation}

We conclude that, in QCG, the criticality parameter ($\Xi_d \neq \Theta_d^{(1)}$) is the one that determines the asymptotic behavior of the mass term in static black holes.

\subsection{Asymptotic form of Weyl tensor and AdS curvature in QCG}\label{Asymptotic Weyl and AdS curvature in QCG}

It is a well-known fact that, in General Relativity, the Weyl tensor adequately packs the components of the Riemann tensor which are not fixed by the Einstein equations. On the other hand, the addition to the Riemann tensor of a precise combination of the Ricci tensor and Ricci scalar acts as a compensator field for conformal transformations of the metric, such that
\begin{equation} \label{Weyl}
W_{\alpha \beta }^{\mu \nu }=R_{\alpha \beta }^{\mu \nu }-\frac{1}{d-1}\,\delta _{[\alpha }^{[\mu }R_{\beta ]}^{\nu ]}+\frac{R}{d(d-1)}\,\delta _{\alpha \beta }^{\mu \nu }\,,
\end{equation}
is a conformally covariant object. Here, the notation $X^{[\mu}Y^{\nu]}=X^{\mu}Y^{\nu}-X^{\nu}Y^{\mu}$ was used. The electric part of the Weyl tensor, defined in eq.\eqref{eq:elWeyl} carries information on gravitational waves propagating in vacuum, tidal forces \cite{Goswami:2019fyk} and, in asymptotically AdS gravity, conserved charges \cite{Ashtekar:1999jx}. Besides, in AdS gravity, there is a tensor which measures the deviation of the spacetimes with respect to a maximally symmetric background (global AdS), known as AdS curvature
\begin{equation}
    F_{\alpha \beta }^{\mu \nu }=R_{\alpha \beta }^{\mu \nu }+\frac{1}{\ell_{0}^{2}}\delta_{\alpha \beta }^{\mu \nu }\,.
\end{equation}
In a Riemannian manifold, the AdS curvature is the only nonvanishing part of the field strength associated to the AdS group.
For Einstein-AdS spaces, the Weyl tensor coincides with the AdS curvature, namely
\begin{equation}
    F_{\alpha \beta }^{\mu \nu }=W_{(E)\alpha \beta }^{\mu \nu }\,.
\end{equation}
In presence of higher-curvature terms, the relation between the on-shell Weyl tensor and the effective AdS tensor
\begin{equation} \label{AdS curvature}
F_{\alpha \beta }^{\mu \nu }=R_{\alpha \beta }^{\mu \nu }+\frac{1}{\ell _{\mathrm{eff}}^{2}}\,\delta _{\alpha \beta }^{\mu \nu }\,,
\end{equation}
is no longer valid.
One may consider the difference between these two notions of curvature
\begin{equation} \label{X def}
X_{\alpha \beta }^{\mu \nu }=W_{\alpha \beta }^{\mu \nu }-F_{\alpha \beta }^{\mu \nu }\,,
\end{equation}
which, by employing the corresponding definitions,  can be evaluated asymptotically for solutions of QCG theory.

By direct use of the definition of the Weyl tensor \eqref{Weyl} and the AdS curvature \eqref{AdS curvature}, this difference is expressed as
\begin{equation}
X_{\alpha \beta }^{\mu \nu }=-\frac{1}{d-1}\,\delta _{\lbrack \alpha }^{[\mu
}R_{\beta ]}^{\nu ]}+\frac{R}{d(d-1)}\,\delta _{\alpha \beta }^{\mu \nu }-
\frac{1}{\ell _{\mathrm{eff}}^{2}}\,\delta _{\alpha \beta }^{\mu \nu }\,.
\end{equation}
The radial expansion of the relevant components of the curvature (see Appendix \ref{ASTAdS}) implies the asymptotic form
\begin{eqnarray}\label{Xtensorcomponents}
X_{tr}^{tr} &=&\frac{(2d-1)p}{r^{2d}}+\mathcal{O}\left( \frac{1}{r^{2d+1}}\right) \,, \notag \\
X_{tn}^{tm} &=&X_{rn}^{rm}=\frac{(d^{2}-3d+1)p}{(d-1)r^{2d}}\,\delta_{n}^{m}+\mathcal{O}\left( \frac{1}{r^{2d+1}}\right)\,,   \notag \\
X_{kl}^{mn} &=&\frac{(3d-1)p}{(d-1)r^{2d}}\,\delta_{kl}^{mn} +\mathcal{O}\left( \frac{1}{r^{2d+1}}\right)\,.
\end{eqnarray}
In sum, the  tensor $X_{\alpha \beta }^{\mu \nu }$ behaves asymptotically as
\begin{equation}
    X_{\alpha \beta }^{\mu \nu }=\mathcal{O}\left( \frac{1}{r^{2d}}\right)\,.
\end{equation}
for the generic metric \eqref{falloff} describing a static black hole in quadratic curvature gravity.
As a result, in many asymptotic computations of physical interest, one may trade off the AdS curvature by the Weyl tensor, without loss of gravitational information.


\section{Kounterterm charges in AAdS Quadratic Curvature Gravity}
\label{kount}

\subsection{Extrinsic counterterms in asymptotically AdS gravity}

The on-shell action of pure Einstein-AdS gravity is proportional to the
volume of the spacetime, which is infinite. Therefore, computations which use this bulk functional to define physical quantities for asymptotically AdS black holes (e.g., mass, angular momentum, free energy, etc.) are expected to be plagued with infrared divergences.
In the context of anti-de Sitter/Conformal Field Theory (AdS/CFT) correspondence \cite{Maldacena:1997re, Gubser:2002tv, Witten:1998qj},
the derivation of holographic correlators at the conformal boundary requires the removal of infinities in the variation of the action, which come from the asymptotic expansion of the canonical momentum \cite{Henningson:1998gx,Skenderis:2002wp}.

In turn, the definition of the canonical momentum in gravity is connected to a well-posed Dirichlet problem for the boundary metric. The appropriate Dirichlet action for
Einstein gravity considers the addition of the Gibbons-Hawking term
\begin{equation}
\tilde{I}=I_{\text{EH}} + \frac{1}{8\pi G} \int\limits_{\partial M} d^{d}x \sqrt{-h}\, K\,,
\end{equation}
which is expressed in terms of the trace of the extrinsic curvature. For a spacetime geometry described by
normal (spatial) coordinates of the form
\begin{equation}
ds^2= N^2(z)dz^2 + h_{ij}(z,x)\,dx^{i}dx^{j}\,,
\end{equation}
the extrinsic curvature adopts a simpler expression respect to the one for a generic ADM (Arnowitt-Deser-Misner) metric, that is,
\begin{equation}
K_{ij}= \frac{1}{2N}\,\partial_{z}\,h_{ij}\,.
\end{equation}

This formula allows to write down an arbitrary variation of the modified action $\tilde{I}$
as
\begin{equation}
\delta\tilde{I}=\int\limits_{M} G^{\mu \nu}\delta g_{\mu \nu} + \int\limits_{\partial M} d^{d}x \frac{1}{2}\sqrt{-h}  \,\pi^{ij}\delta h_{ij}\,,
\end{equation}
where the canonical momentum is
\begin{equation}
\pi^{ij}= \frac{2}{\sqrt{-h}}\frac{\delta\tilde{I}}{\delta h_{ij}}=\frac{1}{8\pi G}\big(K^{ij}-h^{ij}K\big)\,.
\end{equation}

The conservation of the above tensor is a consequence of the Einstein equation $G^{i}_{z}=0$.
However, when used as a Brown-York stress tensor for the theory, this quantity does not lead to finite conserved charges for AAdS spaces. The renormalization of the quasilocal stress tensor and, therefore, the variation of the action, is achieved by adding local counterterms at the boundary. In the standard approach, those counterterms are covariant functions of
the intrinsic/boundary metric as they need to be compatible with the Dirichlet problem for the metric $h_{ij}$ \cite{Balasubramanian:1999re, Emparan:1999pm}.

Within the above framework, a boundary term proportional to the Gibbons-Hawking term with a different overall factor, or a nonlinear contribution in the extrinsic curvature, would necessarily produce
surface terms which contain variations of $K_{ij}$. This situation is analogous to a system in Classical Mechanics where the surface term contains variations of the \emph{velocity}.

It is then surprising that there are cases where the renormalization of both the AdS gravity action and its variation is produced by the addition of extrinsic counterterms. Indeed, this observation can be illustrated by very simple examples:\medskip

\textit{i}) In three spacetime dimensions, the Chern-Simons formulation of AdS gravity gives rise to the
Einstein-Hilbert Lagrangian with negative cosmological constant plus a half of the Gibbons-Hawking term.
While this surface term is clearly at odds with the Dirichlet variational principle for $h_{ij}$, it is
important to notice that it regulates the Euclidean action for black hole solutions \cite{Banados:1998ys}.
The correct use of the asymptotic form of the metric produces the matching with the standard renormalization
prescription \cite{Miskovic:2006tm}.\medskip

\textit{ii}) In four-dimensional AdS gravity, the addition of the Chern form, which is a boundary term nonlinear in the extrinsic curvature, also produces a finite Euclidean action for AAdS spaces \cite{Olea:2005gb}. The  surface term in this case is locally equivalent to a bulk topological (Gauss-Bonnet) term. A puzzling feature
of this renormalized action is the fact that the Dirichlet problem cannot even be defined, as there is no way of getting rid of the variations of the extrinsic curvature. The asymptotic expansion of the total surface
term makes possible to reconcile this extrinsic renormalization with the standard one in the context of gauge/gravity duality \cite{Miskovic:2009bm}.\medskip

The two examples listed above are the simplest cases of a renormalization scheme known as Kounterterms
\cite{Olea:2005gb, Olea:2006vd},
\begin{equation}   I_{\mathrm{ren}}=I_{\mathrm{EH}}+c_{d}\int\limits_{\partial M}d^{d}x\sqrt{-h}\,B_{d}(h, K, \mathcal{R})\,,
\end{equation}
which considers boundary counterterms given as a polynomial of the extrinsic and intrinsic curvatures, that is, $K_{ij}$ and $\mathcal{R}^{i}_{\,jkl}$, respectively. It is simple to verify that this method correctly reproduces the black hole thermodynamics for Schwarzschild-AdS black holes and topological extensions of their cross sections. A less trivial check comes from Kerr-AdS black holes, whose thermal behavior is also appropriately accounted for within this procedure.

The metric for any AAdS spacetime can be written as a power-series expansion in the holographic (radial) coordinate, known as Fefferman-Graham gauge, such that the line element is
\begin{equation}
    ds^2=\frac{\ell^2}{z^{2}}dz^2+\frac{1}{z^{2}}(g_{(0)ij}+z^2g_{(2)ij}+\cdots)\,dx^{i}dx^{j}\,.
\end{equation}

As it is manifest from the asymptotic form of the boundary metric in AAdS gravity,
\begin{equation}
    h_{ij}=\frac{g_{(0)ij}}{z^{2}}+\cdots\,,
\end{equation}
a proper Dirichlet problem for the metric can be defined only for $g_{(0)ij}$ at the conformal boundary \cite{Papadimitriou:2005ii}.

The key observation which allows the addition of Kounterterms in AdS gravity is the fact that the extrinsic
curvature has a similar asymptotic behavior as the one of the boundary metric \cite{Witten:2018lgb}, that is,
\begin{equation}
    K_{ij}=\frac{1}{\ell }\frac{g_{(0)ij}}{z^{2}}+\cdots\,.
\end{equation}

In point of fact, the variation of the extrinsic curvature is also expressed in terms of the
variation of the holographic source $g_{(0)ij}$ and, as a consequence, the addition of extrinsic
counterterms are compatible with a holographic description of AAdS gravity.

Furthermore, the form of the Kounterterms remains the same irrespective of the inclusion of higher-curvature
terms. In doing so, they provide the renormalization of the action in Einstein-Gauss-Bonnet
\cite{Kofinas:2006hr} and Lovelock gravity \cite{Kofinas:2007ns}
with AdS asymptotics. The information on the couplings of the different terms of the theory is somehow
encoded in the overall factor $c_{d}$.

In recent work, this proposal for renormalization of AAdS gravity has been extended to deal with the inclusion of quadratic curvature couplings in the action \cite{Giribet:2018hck, Giribet:2020aks, Miskovic:2022mqv}.
The corresponding renormalized action adopts the form
\begin{equation} \label{IrenQCG}
I_{\mathrm{ren}}=I_{\mathrm{QCG}}+c^{\mathrm{QCG}}_d\int\limits_{\partial M}d^dx\sqrt{-h}\,B_d\,.
\end{equation}

In even bulk dimensions ($d+1=2n$), the Kounterterms are given by
\begin{eqnarray}
&&B_{2n-1}=2n\int\limits^{1}_{0}dt\,\delta^{j_{1}\cdots j_{2n-1}}_{i_{1}\cdots i_{2n-1}}\,K^{i_{1}}_{j_{1}}\left(\frac{1}{2}\,\mathcal{R}^{i_{2}\,i_{3}}_{j_{2}\,j_{3}}-t^{2}K^{i_{2}}_{j_{2}}K^{i_{3}}_{j_{3}}\right)\times\cdots \notag\\
&&\hspace{2.3cm}\cdots\times\left(\frac{1}{2}\,\mathcal{R}^{i_{2n-2}\,i_{2n-1}}_{j_{2n-2}\,j_{2n-1}}-t^{2}K^{i_{2n-2}}_{j_{2n-2}}K^{i_{2n-1}}_{j_{2n-1}}\right)\,,
\end{eqnarray}
whose coupling
\begin{equation}\label{acoplepar}
c_{2n-1}^{\mathrm{QCG}}=-\frac{(-\ell _{\mathrm{eff}}^{2})^{n-1}}{\kappa
n(2n-2)!}\,\left( \rule{0pt}{11pt}1-(2n-1)\omega _{2n-1}\right) \,,
\end{equation}
is chosen by the cancellation of the leading-order divergences in the action.

In turn, the Kounterterms in the action \eqref{IrenQCG} in odd spacetime dimensions ($d+1=2n+1$) are expressed as
\begin{eqnarray}\label{Ktodd}
&&B_{2n}=2n\sqrt{-h}\int\limits_{0}^{1}du\int\limits_{0}^{u}ds\, \delta^{\,j_{1}\cdots j_{2n}}_{\,i_{1}\cdots i_{2n}}K^{i_{1}}_{j_{1}}\delta^{i_{2}}_{j_{2}}\left(\frac{1}{2}\mathcal{R}^{i_{3}\,i_{4}}_{j_{3}\,j_{4}}-u^{2}K^{i_{3}}_{j_{3}}K^{i_{4}}_{j_{4}}+\frac{s^{2}}{\ell^{2}_{\mathrm{eff}}}\delta^{i_{3}}_{j_{3}}\delta^{i_{4}}_{j_{4}} \right)\times\cdots\notag\\
&&\hspace{4cm}\cdots\times\left(\frac{1}{2}\mathcal{R}^{i_{2n-1}\,i_{2n}}_{j_{2n-1}\,j_{2n}}-u^{2}K^{i_{2n-1}}_{j_{2n-1}}K^{i_{2n}}_{j_{2n}}+\frac{s^{2}}{\ell^{2}_{\mathrm{eff}}}\delta^{i_{2n-1}}_{j_{2n-1}}\delta^{i_{2n}}_{j_{2n}} \right),
\end{eqnarray}
with a coupling singled out by the vanishing of the variation of the total action at leading order, that is,
\begin{equation} \label{acopleQCG}
c_{2n}^{\mathrm{QCG}}=\frac{(-\ell _{\mathrm{eff}}^{2})^{n-1}}{%
2^{2n-2}\kappa n(n-1)!^{2}}\,\left( \rule{0pt}{11pt}1-2n\omega _{2n}\right)
\,.
\end{equation}

\subsection{Conserved quantities}

For an arbitrary gravity theory, with a Lagrangian  $\mathcal{L}=\mathcal{L}(g,R)$, the Noether-Wald procedure leads to a conserved charge which stems from diffeomorphic invariance of the bulk action, without additional boundary terms \cite{Iyer:1994ys, Anastasiou:2017rjf}.
The Noether charge is obtained from the prepotential
\begin{equation}
 Q^{\mu \nu }= 2\left(E^{\alpha \beta}_{\mu \sigma}\nabla^{\mu}\xi^{\sigma}+2\xi^{\sigma}\nabla^{\mu}E^{\alpha \beta}_{\mu \sigma} \right)\,,
\end{equation}
in terms of the tensor $E^{\alpha \beta}_{\mu \nu}= \frac{\partial \mathcal{L}}{\partial R^{\mu \nu}_{\alpha \beta}}$, and expressed as an integral on a co-dimension 2 surface $\Sigma$
\begin{equation}
Q[\xi ]=\int\limits_{\Sigma }d\Sigma _{\mu \nu }Q^{\mu \nu }\,,
\end{equation}
where  $d\Sigma
_{\mu \nu }=\frac{1}{2}\,d^{d-1}x\sqrt{\sigma }\,(n_{\mu }u_{\nu }-n_{\nu }u_{\mu })$ is the
covariant surface element. The normals $n_{\mu }$ and $u_{\mu }$ are spacelike and timelike unit vectors, respectively, orthogonal to the transversal cross section.

Boundary terms clearly modify the above definition of conserved quantities. Indeed, the addition of Kounterterms renormalizes the prepotential, which acquires the form
\begin{equation}
    Q^{\mu \nu}_{\mathrm{ren}}=Q^{\mu \nu}+c_{d}B^{[\mu}\xi^{\nu]}\,,
\end{equation}
where the boundary term can be trivially extended to a bulk vector as $B_{d}=n_{\mu}B^{\mu}$.

Consider a spacetime whose metric is written in Schwarzschild-like coordinates, $x^{\mu}=(r, x^{i})$, given by the line element
\begin{equation}\label{normalSch}
ds^{2}=\mathcal{N}^{2}(r)dr^{2}+h_{i j}(r,x)dx^{i}dx^{j}\,,
\end{equation}
where $r$ is a radial coordinate and $h_{i j}$ is the induced metric on the boundary, defined at a constant radius. This foliation is generated by a normal vector $n_{\mu}n^{\mu}=1$, such that $n_{\mu}=(n_{r}, n_{i})=(\mathcal{N}, \vec{0})$.

On the other hand, Kounterterms are mathematical structures constructed out from the boundary geometry. Only in even spacetime dimensions, they can be related to bulk (topological) invariants, what produces fully covariant Noether-Wald charges \cite{Giribet:2018hck, Giribet:2020aks}.

In order to present a unified discussion for even and odd dimensions, we consider the projection of the Noether prepotential
\begin{equation}
n_{\nu }Q^{i\nu }=q_{j}^{i}\,\xi ^{j}\,,  \label{comparedcharge}
\end{equation}
such that the charge density tensor $q_{j}^{i}$ is
introduced.
This definition casts the Noether charge in a closer form respect to the one obtained by quasilocal methods \cite{Emparan:1999pm}
\begin{eqnarray}\label{Q}
    Q[\xi]&=&\int\limits_{\Sigma} d^{d-1}x\,\sqrt{\sigma}u_i\,q^{i}_j\, \xi^j\,.
\end{eqnarray}
The above formula assumes a topology of the boundary $\partial\mathcal{M}=\mathbb{R}\times\Sigma$, such that the induced metric can be written in an ADM coordinate frame
\begin{equation}
   h_{i j}\, dx^{i}dx^{j}=-\tilde{N}^{2}dt^{2}+\sigma_{m n}(dy^{m}+N^{m}dt)(dy^{n}+N^{n}dt)\,,
\end{equation}
where $\sigma_{n m}$ is the metric on $\Sigma$ and $x^{i}=(t, y^{m})$ are local coordinates on the boundary. The unit vector $u_{i}u^{i}=\mb{-}1$, normal to the surface $\Sigma$ is given by $u_{i}=(u_{t}, u_{m})=(-\tilde{N}, \vec{0})$.\\

For AAdS spacetimes in even-dimensional QCG, conserved quantities are measured by the charge density
\begin{eqnarray}\label{chargex1}
    q^{i}_{j}&=&\frac{1}{ (2n-2)!2^{n-2}}\delta^{i i_{2}\cdots i_{2n-1}}_{j_{1} j_{2}\cdots j_{2n-1}}K^{j_{1}}_{j}\left[nc^{\mathrm{QCG}}_{2n-1}(2n-2)!R^{j_{2} j_{3}}_{i_{2} i_{3}}\cdots R^{j_{2n-2} j_{2n-1}}_{i_{2n-2} i_{2n-1}}+\right.\notag\\
    &+&\left.\left(\frac{1}{\kappa}\delta^{j_{2} j_{3}}_{i_{2} i_{3}}+\alpha R^{r}_{r}\delta^{j_{2} j_{3}}_{i_{2} i_{3}}+2\beta R\delta^{j_{2} j_{3}}_{i_{2} i_{3}}+2\gamma(2n-2)(2n-3)R^{j_{2} j_{3}}_{i_{2} i_{3}}\right)\delta^{j_{4} j_{4}}_{i_{4} i_{5}}\cdots\delta^{j_{2n-2} j_{2n-1}}_{i_{2n-2} i_{2n-1}} \right]\notag\\
    &-&2N\left[\alpha \left(\nabla^{r}R^{i}_{j}- \nabla^{i}R^{r}_{j}+ \nabla^{r}R^{r}_{r}\delta^{i}_{j}+ \nabla^{l}R^{r}_{l}\delta^{i}_{j}\right)+2\beta \nabla^{r}R\delta^{i}_{j}\right]-\alpha N\left(\nabla^{k}R_{k}^{r}\delta^{i}_{j}-\nabla^{i}R^{r}_{j}\right)\,\notag\\
    &+&2\alpha K^{l}_{j}R^{i}_{l}\,.
\end{eqnarray}
 By an abuse of notation, the extrinsic curvature in Schwarzschild-like coordinates \eqref{normalSch} is defined by   $K_{i j}=-(1/2\mathcal{N})\,\partial_{r}h_{i j}\,$.

The presence of covariant derivatives in the curvature is a consequence of the higher-derivative nature of QCG theory. Therefore, it generalizes the notion of energy via Kounterterms beyond the restriction imposed by Lovelock theorem \cite{Lovelock:1971yv}.\\

In the odd-dimensional case,
the corresponding charge density tensor appears modified as $q_{i}^{j}+q_{(0)i}^{j}$, due to the existence of the vacuum energy $Q_{(0)}[\xi]$. The mass of the gravitational solutions, $Q[\xi]$, is then given by the first term, with a charge tensor \cite{Miskovic:2022mqv}
\begin{eqnarray}
\label{chargeoddQCG}
q_{j}^{i} &=&\frac{1}{2^{n-2}}\,\delta _{kj_{1}\cdots
j_{2n-1}}^{ii_{1}\cdots i_{2n-1}}\,K_{j}^{k}\delta _{i_{1}}^{j_{1}}\left[
nc_{2n}^{\mathrm{QCG}}\int\limits_{0}^{1}du\left(
R_{i_{2}i_{3}}^{j_{2}j_{3}}+\frac{u^{2}}{\ell _{\mathrm{eff}}^{2}}\,\delta
_{i_{2}i_{3}}^{j_{2}j_{3}}\right) \cdots \left(
R_{i_{2n-2}i_{2n-1}}^{j_{2n-2}j_{2n-1}}+\frac{u^{2}}{\ell _{\mathrm{eff}}^{2}}\,\delta _{i_{2n-2}i_{2n-1}}^{j_{2n-2}j_{2n-1}}\right) \right.   \notag \\
&+&\left. \frac{1}{(2n-1)!}\left( \frac{1}{\kappa }\,\delta
_{i_{2}i_{3}}^{j_{2}j_{3}}+\alpha \,R_{r}^{r}\delta
_{i_{2}i_{3}}^{j_{2}j_{3}}+2\beta \,R\delta
_{i_{2}i_{3}}^{j_{2}j_{3}}+2(2n-1)(2n-2)\gamma
R_{i_{2}i_{3}}^{j_{2}j_{3}}\right) \delta _{i_{4}i_{5}}^{j_{4}j_{5}}\cdots
\delta _{i_{2n-2}i_{2n-1}}^{j_{2n-2}j_{2n-1}}\rule{0pt}{20pt}\right]   \notag
\\
&&-N\alpha \left[ \rule{0pt}{15pt}2 \left( \nabla ^{r}R_{j}^{i}-\nabla
^{i}R_{j}^{r}+\nabla ^{r}R_{r}^{r}\delta _{j}^{i}+\nabla ^{k}R_{k}^{r}\delta
_{j}^{i}\right) +\nabla ^{k}R_{k}^{r}\delta _{j}^{i}-\nabla ^{i}R_{j}^{r}%
\right]   \notag \\
&&+\, 2\alpha \,K_{k}^{i}R_{j}^{k}-4N\beta \,\nabla ^{r}R\,\delta _{j}^{i}\,,
\end{eqnarray}
while the vacuum energy is produced by the tensor
\begin{eqnarray}
q_{(0)j}^{i} &=&nc^{\mathrm{QCG}}_{2n}\,\delta _{k j_{1}\cdots j_{2n-1}}^{i i_{1}\cdots i_{2n-1}}\int\limits_{0}^{1}du\,u\left(
K_{j}^{k}\delta _{i_{1}}^{j_{1}}+K_{i_{1}}^{k}\delta _{j}^{j_{1}}\right) \left( \frac{1}{2}\,\mathcal{R}_{i_{2}i_{3}}^{j_{2}j_{3}}-u^{2}K_{i_{2}}^{j_{2}}K_{i_{3}}^{j_{3}}+\frac{u^{2}}{\ell _{\mathrm{eff}}^{2}}\,\delta _{i_{2}}^{j_{2}}\delta_{i_{3}}^{j_{3}}\right) \times   \notag \\
&&\qquad \cdots \times \left(  \frac{1}{2}\,\mathcal{R}_{i_{2n-2}\,i_{2n-1}}^{j_{2n-2}\,j_{2n-1}}-u^{2}K_{i_{2n-2}}^{j_{2n-2}}K_{i_{2n-1}}^{j_{2n-1}}+\frac{u^{2}}{\ell _{\mathrm{eff}}^{2}}\,\delta _{i_{2n-2}}^{j_{2n-2}}\delta_{i_{2n-1}}^{j_{2n-1}}\right) \,.  \label{chargevacuumQCG}
\end{eqnarray}
Notice that, as it evident from the appearance of the coupling $c_{2n}^{\mathrm{QCG}}$ as an overall factor in the last expression, the vacuum energy is an effect of the addition of regulating counterterms. No further analysis on the properties of the vacuum energy tensor \eqref{chargevacuumQCG} will be performed here, since the focus  is on the black hole mass and its link to the electric part of the Weyl tensor.


\section{Conformal Mass in QCG }
\label{conf}

In asymptotically AdS spaces, the form of the extrinsic curvature at large distances is generically given by
\begin{equation}\label{KAAdS}
    K^{i}_{j}=-\frac{1}{\ell_{\mathrm{eff}}}\,\delta^{i}_{j}+\mathcal{O}\left(\frac{1}{r^{2}}\right)\,,
\end{equation}
such that the Ricci tensor, for static black holes (see Appendix \ref{ASTAdS}), behaves as \begin{equation}\label{RicciAAdS}
   R^{\mu}_{\nu}=-\frac{d}{\ell^{2}_{\mathrm{eff}}}\,\delta^{\mu}_{\nu}+\mathcal{O}\left(\frac{1}{r^{2d}}\right) \,.
\end{equation}

In this section, the aim is to prove that, without loss of information on the conserved quantities, the charge tensors in eqs.\eqref{chargex1} and \eqref{chargeoddQCG} can be consistently truncated as the electric part of the Weyl tensor.
To this end, the strategy adopted is as follows:
The use of this asymptotic falloff in the curvature allows to simplify the expression in the charge tensor. Indeed,  the covariant derivatives acting on curvature tensors can be dropped from \eqref{chargex1}  and \eqref{chargeoddQCG}.
The Kounterterms --and the charges derived from them-- adopt a quite different form in the even and odd-dimensional cases. Despite this fact, it can be shown that the charge density is expressed in terms of the AdS curvature as
\begin{equation}\label{factorizeddensitycharge}
    q^{i}_{j}=a_{d}~\delta_{j_{1}j_{2}\cdots j_{d}}^{\,i\; i_{2}\,\cdots
i_{d}}K^{j_{1}}_{j}F_{i_{2}\,i_{3}}^{j_{2}j_{3}}\mathcal{J}_{\,i_{4}\,\cdots
i_{d}}^{j_{4}\cdots j_{d}}(F)\,,
\end{equation}
irrespective of the spacetime dimension. Here, $a_{d}$ is coupling-dependent constant, which also depends on the effective AdS radius, and $\mathcal{J}(F)$ is a polynomial of the AdS curvature, whose leading order is $\mathcal{O}(1)$. Since the AdS curvature vanishes for global AdS space ($F_{\alpha
\beta }^{\mu \nu }=0$), and the fact that $\mathcal{J}(0)$ is finite, the Noether charge is identically zero, as expected for the vacuum state of the theory.

Working out the asymptotic behavior of the different tensors in the charge formula \eqref{Q}, one may notice that $u_{t}=-\sqrt{-g_{tt}}=%
\mathcal{O}(r)$ and $\sqrt{\sigma }=\mathcal{O}(r^{d-1})$. As a consequence, a finite contribution in
the charge only appears if $q_{j}^{i}$ is that of order $\mathcal{O}(r^{-d})$. This also justifies the fact that the extrinsic curvature can be truncated as the leading order in eq.\eqref{KAAdS}, i.e., $K^i_j=-1/\ell_{\mathrm{eff}}$. Then, the AdS curvature can be consistently traded off by the Weyl tensor since $F=W+\mathcal{O}\left( r^{-2d}\right)$ as pointed out in Sec.~\eqref{Asymptotic Weyl and AdS curvature in QCG}. The previous analysis implies that the charge density tensor \eqref{factorizeddensitycharge} can be written as
\begin{equation}
    q^{i}_{j}=-\frac{a_{d}}{\ell_{\mathrm{eff}}}~\delta_{j_{1}j_{2}\cdots j_{d}}^{\,i\; i_{2}\,\cdots
i_{d}}\delta^{j_{1}}_{j}W_{i_{2}\,i_{3}}^{j_{2}j_{3}}\mathcal{J}_{\,i_{4}\,\cdots
i_{d}}^{j_{4}\cdots j_{d}}(F)\,.
\end{equation}

Another key ingredient in the derivation is the asymptotic form of the polynomial $\mathcal{J}(F)$. It can be shown that only the finite part of polynomial plays a role in the evaluation of the conserved quantities. This leading-order term is dictated by the asymptotic form of the curvature, what makes no difference between global AdS and black hole spacetimes. Therefore, one may approximate the charge by setting $F=0$, resulting in
\begin{equation}
    q^{i}_{j}=-\frac{a_{d}}{\ell_{\mathrm{eff}}}~\delta_{j_{1}j_{2}\cdots j_{d}}^{\,i\; i_{2}\,\cdots
i_{d}}\delta^{j_{1}}_{j}W_{i_{2}\,i_{3}}^{j_{2}j_{3}}\mathcal{J}_{\,i_{4}\,\cdots
i_{d}}^{j_{4}\cdots j_{d}}(0)+\mathcal{O}\left( \frac{1}{r^{2d}}\right)\,,
\end{equation}
without loss of information about the energy of the system.

In what follows, the extensive use of the fall-off of the fields involved, and yet another power-counting argument, would lead to a charge density proportional to the electric part of the Weyl tensor. In this way, the AMD definition of asymptotic charges can be extended to QCG.

As a last step, using the definition of the electric part of the Weyl tensor \eqref{eq:elWeyl}, we obtain that the final result of the leading-order term of the charge density tensor in QCG, in terms of the electric part of the Weyl tensor, is
\begin{equation}
q_{j}^{i}=-\frac{2\ell _{\mathrm{eff}}\,\Xi _{d}}{\kappa }\,\mathcal{E}%
_{j}^{i}+\mathcal{O}\left(\frac{1}{r^{2d}}\right) \,.  \label{q density}
\end{equation}%
 Finally, the conserved charge in the case of QCG corresponds to
\begin{equation}\label{QQCG}
    Q_{\mathrm{QCG}}[\xi]=-\frac{2\ell _{\mathrm{eff}}\,\Xi _{d}}{\kappa}\int\limits_{\Sigma}d^{d-1}x\sqrt{\sigma}\,u_{i}\,\mathcal{E}_{j}^{i}\,\xi^{j}\,.
\end{equation}

\subsection{Even dimensions}

In the even-dimensional case, upon imposing the condition on the curvature for AAdS spaces \eqref{RicciAAdS} in the charge density \eqref{chargex1}, it takes the form
\begin{eqnarray}\label{evenchargeasymp}
q_{j}^{i} &=&\frac{1}{(2n-2)!2^{n-2}}\,\delta _{j_{1}j_{2}\cdots
j_{2n-1}}^{ii_{2}\cdots i_{2n-1}}K_{j}^{j_{1}}\left[ \rule{0pt}{17pt}nc_{2n-1}^{\mathrm{QCG}}(2n-2)!\,R_{i_{2}i_{3}}^{j_{2}j_{3}}\cdots
R_{i_{2n-2}i_{2n-1}}^{j_{2n-2}j_{2n-1}}\right. \notag \\
&+&\left( \frac{1}{\kappa }-\alpha \,\frac{2(2n-1)}{\ell _{\mathrm{eff}}^{2}}-\beta \,\frac{4n(2n-1)}{\ell _{\mathrm{eff}}^{2}}\right) \delta _{i_{2}i_{3}}^{j_{2}j_{3}}\delta
_{i_{4}i_{5}}^{j_{4}j_{5}}\cdots \delta
_{i_{2n-2}i_{2n-1}}^{j_{2n-2}j_{2n-1}}\notag\\
&+&\left.2\gamma (2n-2)(2n-3)R_{i_{2}i_{3}}^{j_{2}j_{3}}\delta
_{i_{4}i_{5}}^{j_{4}j_{5}}\cdots \delta
_{i_{2n-2}i_{2n-1}}^{j_{2n-2}j_{2n-1}}\rule{0pt}{17pt} \right]+\mathcal{O}\left(\frac{1}{r^{4n-2}}\right)\,.
\end{eqnarray}

In this respect, it is particularly convenient to define the following constants,
\begin{eqnarray}
A &=&\frac{1}{\kappa }-\frac{2\alpha (2n-1)}{\ell _{\mathrm{eff}}^{2}}-\frac{4n\beta (2n-1)}{\ell _{\mathrm{eff}}^{2}}\,,  \notag \\
C &=&\frac{2\gamma (2n-2)(2n-3)}{\ell _{\mathrm{eff}}^{2}}\,,
\end{eqnarray}
such that the coupling of the boundary term \eqref{acoplepar} becomes
\begin{equation}
c_{2n-1}^{\mathrm{QCG}}=\frac{(-\ell _{\mathrm{eff}}^{2})^{n-1}}{n(2n-2)!}\,(C-A)\,.
\end{equation}%
Equipped with these redefinitions of constants, the charge can be recast as
\begin{eqnarray}\label{evencharge3}
q_{j}^{i} &=&\frac{1}{(2n-2)!2^{n-2}}\,\delta _{j_{1}j_{2}\cdots
j_{2n-1}}^{ii_{2}\cdots i_{2n-1}}K_{j}^{j_{1}}\left[(-\ell _{\mathrm{eff}}^{2})^{n-1}\,(C-A)\,R_{i_{2}i_{3}}^{j_{2}j_{3}}\cdots
R_{i_{2n-2}i_{2n-1}}^{j_{2n-2}j_{2n-1}}\right. \notag\\
&+&\left.\!\!A\, \delta _{i_{2}i_{3}}^{j_{2}j_{3}}\delta
_{i_{4}i_{5}}^{j_{4}j_{5}}\cdots \delta
_{i_{2n-2}i_{2n-1}}^{j_{2n-2}j_{2n-1}}+C\ell^{2}_{\mathrm{eff}}R_{i_{2}i_{3}}^{j_{2}j_{3}}\delta
_{i_{4}i_{5}}^{j_{4}j_{5}}\cdots \delta
_{i_{2n-2}i_{2n-1}}^{j_{2n-2}j_{2n-1}}\right]\!+\mathcal{O}\left(\frac{1}{r^{4n-2}}\right).
\end{eqnarray}
In order to streamline the discussion, the following shorthand notation is introduced
\begin{equation}
\begin{array}[b]{llll}
\delta _{i_{1}\cdots i_{p}}^{j_{1}\cdots j_{p}} & \rightarrow \quad \delta
^{^{[p]}}\,,\hspace{1cm} & R_{i_{1}\,i_{2}}^{j_{1}\,j_{2}} &
\rightarrow \quad R\,,\medskip  \\
\dfrac{1}{\ell _{\mathrm{eff}}^{2}}\,\delta _{i_{1}\,i_{2}}^{j_{1}\,j_{2}} &
\rightarrow \quad \Delta \,, & K_{j}^{i} & \rightarrow \quad K\,,
\end{array}
\end{equation}
such that contracted indices are omitted.
With the use of this notation, the expression \eqref{evencharge3} can be further simplified as
\begin{equation}\label{chargex4}
q_{j}^{i}=\frac{(-\ell _{\mathrm{eff}}^{2})^{n-1}}{(2n-2)!2^{n-2}}\,\delta
^{i[2n-1]}K_{j}\left[ A\left( \rule{0pt}{13pt}(-\Delta
)^{n-1}-R^{n-1}\right) +CR\left( \rule{0pt}{13pt}R^{n-2}-\left( -\Delta
\right) ^{n-2}\right) \right] .
\end{equation}
The above formula is such that one can apply the identity \eqref{b-a}, with $b=R$ and $a=-\Delta $. A proper factorization of this charge produces an expression proportional to the AdS curvature $F=R+\Delta $, which can be written as
\begin{equation}\label{TensorialFactEvenCharge}
q_{j}^{i}=\frac{(-1)^{n-1}\ell _{\mathrm{eff}}^{2n-2}}{2^{n-2}(2n-2)!}\,\delta_{k\,j_{2}\cdots j_{2n-1}}^{ii_{2}\cdots
i_{2n-1}}K_{j}^{k}F_{i_{2}i_{3}}^{j_{2}j_{3}}\mathcal{J}_{i_{4}\cdots
i_{2n-1}}^{j_{4}\cdots j_{2n-1}}(F)\,.
\end{equation}
Here,  the tensor $\mathcal{J}_{i_{4}\cdots i_{2n-1}}^{j_{4}\cdots j_{2n-1}}(F)\equiv\mathcal{J}(F)$ is a totally antisymmetric tensor in upper (lower)
indices, constructed as the product of curvatures and Kronecker deltas of rank 2, that is,
\begin{equation}
\mathcal{J}(F)=\int\limits_{0}^{1}du\,\left[ \rule{0pt}{12pt} C(n-2)(F-\Delta) (uF-\Delta) ^{n-3}-A(n-1) (uF-\Delta)^{n-2}\right] \,.  \label{Jtensor}
\end{equation}

As argued previously, the only relevant contribution from the tensor \eqref{Jtensor} is its finite part, which is equivalent to evaluating the above expression for $F=0$. Then, the integration in the parameter $u$ can be trivially performed, to get
\begin{eqnarray}
\mathcal{J}_{\,\,\,i_{4}\cdots i_{2n-1}}^{j_{4}\cdots j_{2n-1}}(0)&=&\left( -1 \right)^{n-2}\left[ \rule{0pt}{12pt}C(n-2)-A\left( n-1\right) \right]
\Delta_{i_{4}i_{5}}^{j_{4}j_{5}} \cdots \Delta_{i_{2n-2}i_{2n-1}}^{j_{2n-2}j_{2n-1}}\,,\notag\\
&=&\left( -1 \right)^{n-2}\Xi_{2n-1} \frac{\left(
n-1\right)}{\kappa \left(\ell _{\mathrm{eff}}^{2}\right) ^{n-2}}%
\, \delta_{i_{4}i_{5}}^{j_{4}j_{5}} \cdots \delta_{i_{2n-2}i_{2n-1}}^{j_{2n-2}j_{2n-1}}\,,
\end{eqnarray}
where the criticality function is given by eq.\eqref{Xi_d},
\begin{equation}
\Xi _{2n-1}=1-(2n-1)\omega _{2n-1}+\frac{4(2n-4)}{\ell _{\mathrm{eff}}^{2}}%
\,\kappa \gamma \,.
\end{equation}
As a consequence, the charge density can be consistently truncated as an expression linear in the Weyl tensor, as the fall-off of the AdS curvature is prescribed by the discussion in Sec.~\ref{Asymptotic Weyl and AdS curvature in QCG}. Then, the relevant tensor reduces to the equation
\begin{equation}
q_{j}^{i}=-\frac{\ell _{\mathrm{eff}}\Xi_{2n-1} }{2\kappa (2n-3)}\,\delta _{jj_{2}j_{3}}^{ii_{2}i_{3}}W_{i_{2}i_{3}}^{j_{2}j_{3}}+\mathcal{O}\left(\frac{1}{r^{4n-2}}\right)\,,
\end{equation}
what can be re-expressed as
\begin{equation}
q_{j}^{i}=\frac{2\ell _{\mathrm{eff}}\Xi_{2n-1} }{\kappa (2n-3)}\,W_{j\ell}^{i\ell}+\mathcal{O}\left(\frac{1}{r^{4n-2}}\right)\,,
\end{equation}
due to the fact the double subtrace (on the boundary indices) of the bulk Weyl tensor is zero. The tracelessness of the Weyl tensor also implies
\begin{equation}
    W_{j\ell}^{i\ell}=-W_{jr}^{ir}=-n_{\mu}n^{\nu}W_{j\nu}^{i\mu}\,,
\end{equation}
such that the charge density can be cast in the form
\begin{equation}
q_{j}^{i}=-\Xi_{2n-1}\,\frac{2\ell _{\mathrm{eff}} }{\kappa (2n-3)}\,    n_{\mu}n^{\nu}W_{j\nu}^{i\mu}\,.
\end{equation}
 Finally, as anticipated in eq.\eqref{QQCG}, the notion of Conformal Mass in QCG theory is obtained as the surface integral
\begin{equation}
    {\cal{H}}_{\mathrm{QCG}}[\xi]=-\Xi _{2n-1}\frac{2\ell _{\mathrm{eff}}\,}{\kappa }\int\limits_{\Sigma}d^{2n-2}x\sqrt{\sigma}\,u_{i}\,\mathcal{E}%
_{j}^{i}\,\xi^{j}\,.
\end{equation}


\subsection{Odd dimensions}

In the central aspects, the  derivation of Conformal Mass in odd-dimensional QCG does not differ much respect to the one in even dimensions. As a matter of fact,  the asymptotic form of the curvature \eqref{RicciAAdS} plugged in the general expression  \eqref{chargeoddQCG} leads to a charge density tensor
\begin{eqnarray}\label{oddchargeasymp}
q^{i}_{j} &=&\frac{1}{2^{n-2}}\,\delta _{k j_{1}\cdots
j_{2n-1}}^{i i_{1}\cdots i_{2n-1}}K_{j}^{k}\delta _{j_{1}}^{i_{1}} \left[ nc_{2n}^{\mathrm{QCG}}\int\limits_{0}^{1}du\left(\! R^{j_{2}j_{3}}_{i_{2}i_{3}}+\frac{u^{2}}{\ell _{\mathrm{eff}}^{2}}\,\delta ^{j_{2}j_{3}}_{i_{2}i_{3}}\!\right) \!\cdots\!
\left( \!R^{j_{2n-2}j_{2n-1}}_{i_{2n-2}i_{2n-1}}+\frac{u^{2}}{\ell _{\mathrm{eff}}^{2}}\,\delta^{j_{2n-2}\,j_{2n-1}}_{i_{2n-2}\,i_{2n-1}}\!\right)\right.\notag \\
&+&\!\!\left. \frac{1}{(2n-1)!}\left( \frac{1}{\kappa }-\alpha \,\frac{4n}{\ell _{\mathrm{eff}}^{2}}-\beta
\,\frac{4n\left( 2n+1\right) }{\ell _{\mathrm{eff}}^{2}}\,\right)\delta_{i_{2} i_{3}}^{j_{2} j_{3}}\delta ^{j_{4}j_{5}}_{i_{4}i_{5}}\cdots
\delta^{j_{2n-2}\,j_{2n-1}}_{i_{2n-2}\,i_{2n-1}}\right.\notag\\
&+&\left.\gamma\, \frac{2(2n-1)(2n-2)}{(2n-1)!}R_{i_{2} i_{3}}^{j_{2} j_{3}}\delta ^{j_{4}j_{5}}_{i_{4}i_{5}}\cdots
\delta^{j_{2n-2}\,j_{2n-1}}_{i_{2n-2}\,i_{2n-1}}
\right]+\mathcal{O}\left(\frac{1}{r^{4n}}\right).\rule{0pt}{20pt}
\end{eqnarray}
In what follows, it is convenient to rewrite the coupling constant \eqref{acopleQCG} using the parametric integration in eq.\eqref{parametric}, such that it adopts the form
\begin{equation}
c_{2n}^{\mathrm{QCG}}=-\frac{\ell _{\mathrm{eff}}^{2(n-1)}}{n\kappa (2n-1)!}\,(1-2n\omega _{2n})\left[ \int\limits_{0}^{1}du\,\left( u^{2}-1\right) ^{n-1}\right]
^{-1},  \label{acopleQCGparametric}
\end{equation}
The integral representation of the coupling $c_{2n}^{\mathrm{QCG}}$ for the Kounterterms in this case leads to an integral form for the charge \eqref{acopleQCGparametric}, that is,
\begin{equation}
q^{i}_{j}=\frac{nc_{2n}^{\mathrm{QCG}}}{2^{n-2}}\,\delta_{k}^{i[2n-1]}\,K^{k}_{j}\int\limits_{0}^{1}du\,\mathcal{Q}(u, F)\,,
\label{modificadocharga2}
\end{equation}
with the corresponding polynomial given by
\begin{eqnarray}\label{Qtensor}
\mathcal{Q}(u, F) &=&\left[ \left( \rule{0pt}{13pt}F+(u^{2}-1)\Delta \right) ^{n-1}-\left( \rule{0pt}{13pt}
(u^{2}-1)\Delta \right) ^{n-1}\right]   \notag \\
&&-\frac{2(2n-1)(2n-2)\kappa \gamma }{\ell _{\mathrm{eff}}^{2}}\,F\,\left(
u^{2}-1\right) ^{n-1}\Delta ^{n-2}\,,
\end{eqnarray}
using the shorthand notation defined in the previous subsection and the AdS curvature $F=R+\Delta$.

The latter expression can be recast into a more convenient form, employing yet another parametric integral. In order to do so, the integral identity \eqref{b-a} can be applied on the first line of eq.\eqref{Qtensor}, with a continuous parameter $s$ instead of $u$, the power $p=n-1$, and the corresponding factors $b=F+(u^{2}-1)\Delta $ and $a=(u^{2}-1)\Delta $, what results in
\begin{equation}
\left( \rule{0pt}{13pt}F+(u^{2}-1)\Delta \right) ^{n-1}-\left(\rule{0pt}{13pt}(u^{2}-1)\Delta \right)
^{n-1}=(n-1)F\int\limits_{0}^{1}ds\,\left( \rule{0pt}{13pt} sF+(u^{2}-1)\Delta \right) ^{n-2}\,.
\end{equation}
Making explicit the integration in $s$ has the clear advantage that the curvature $F$ can be readily factorized from the $\mathcal{Q}$ tensor, such that \eqref{oddchargeasymp} becomes
\begin{equation}\label{TensorialFactOddCharge}
q^{i}_{j}=\frac{(-1)^{n}\ell _{\mathrm{eff}}^{2n-2} }{2^{3n-4}\kappa (n-1)!^{2}}\,\delta_{k j_{2} \cdots j_{2n}}^{i\, i_{2} \cdots i_{2n}}K_{j}^{k} F^{j_{2} j_{3}}_{i_{2} i_{3}}\mathcal{J}^{j_{4} \cdots j_{2n}}_{i_{4} \cdots i_{2n}}(F)\,,
\end{equation}
with a polynomial $\mathcal{J}(F)$ in the same fashion as in the even-dimensional case,
\begin{equation}
    \mathcal{J}(F)=(n-1)\int\limits_{0}^{1}du\,\left[ (1-2n\omega_{2n})\int\limits_{0}^{1}ds\,\left( \rule{0pt}{13pt}sF+(u^{2}-1)\Delta \right)^{n-2}-\frac{4(2n-1)\kappa \gamma }{\ell _{\mathrm{eff}}^{2}}\,\left(
u^{2}-1\right) ^{n-1}\Delta ^{n-2}\right]\,.
\end{equation}
Because of a series of arguments given above, the physical information on the energy of the system is encoded in the AdS curvature $F$, what renders irrelevant the subleading contributions in both the extrinsic curvature and the polynomial $\mathcal{J}(F)$. Therefore, it suffices to evaluate the polynomial for $F=0$, what leaves its finite value
\begin{equation}
\mathcal{J}(0)=(n-1)\int\limits_{0}^{1}du\,\left[ (1-2n\omega_{2n})(u^{2}-1)^{n-2}-\frac{4(2n-1)\kappa \gamma }{\ell_{\mathrm{eff}}^{2}}\,\left( u^{2}-1\right) ^{n-1}\right] \Delta ^{n-2}.
\end{equation}
The identity \eqref{u^2-1} is of particular use to leave this expression in the form
\begin{equation} \label{J(0)'}
\mathcal{J}(0)=\left( 2n\omega_{2n}-1-\frac{8(n-1)}{\ell _{\mathrm{eff}}^{2}}\,\kappa \gamma\right) \frac{2n-1}{2}\,\Delta
^{n-2}\int\limits_{0}^{1}du\,\left( u^{2}-1\right) ^{n-1}\,,
\end{equation}
where the parametric integration can be trivially performed, such that
\begin{equation}
\mathcal{J}(0)=\Xi_{2n}\,\frac{(-1)^{n}2^{2n-3}(n-1)!^{2}\ell^{2n-4}_{\mathrm{eff}}}{(2n-2)!}\,(\delta^{[2]})^{n-2}\,,
\end{equation}
as a consequence of the relation that follows from eq.\eqref{Xi_d},
\begin{equation}
2n\omega_{2n}-1-\frac{8(n-1) }{\ell _{\mathrm{eff}}^{2}}\,\kappa \gamma =-\Xi _{2n}\,.
\end{equation}

This reasoning, added to the corresponding falloff of the fields and the relation between the AdS curvature and the Weyl tensor in Sec.\eqref{Asymptotic Weyl and AdS curvature in QCG}, leads to a charge density tensor
\begin{equation}
    q^{i}_{j}=-\frac{\ell_{\mathrm{eff}}\Xi_{2n}}{2\kappa(2n-2)}\delta
_{j j_{2} j_{3}}^{i\,i_{2}\,i_{3}}W_{i_{2}\,i_{3}}^{j_{2}j_{3}}+\mathcal{O}\left(\frac{1}{r^{4n}}\right)\,,
\end{equation}
where the extrinsic curvature has been replaced by its leading order in eq.\eqref{KAAdS}.
From this point, the final part of the derivation of AMD charges is analog to the one of the even-dimensional case. The formula is then proportional to the criticality condition, also in odd-dimensional QCG theory, that is,
\begin{equation}\label{AMDodd}
    {\cal{H}}_{\mathrm{QCG}}[\xi]=-\Xi _{2n}\frac{2\ell _{\mathrm{eff}}\,}{\kappa }\int\limits_{\Sigma}d^{2n-1}x\sqrt{\sigma}\,u_{i}\,\mathcal{E}%
_{j}^{i}\,\xi^{j}\,.
\end{equation}


\subsection{Mass for static black holes in QCG}

In order to test the expression for the Conformal Mass in QCG \eqref{QQCG}, the computation of the energy of a static black hole \eqref{ansatzTBH} can be performed. This solution has an obvious isometry that corresponds to the time translation, given in terms of a Killing vector $\xi^{i}=\delta^{i}_{t}$.

The formula for the conserved quantities requires a unit vector $u_{i}=-f\delta^{t}_{i}$ and the determinant of the metric in the codimension-2 surface $\Sigma$, that is, $\sqrt{\sigma}=r^{d-1}\sqrt{\gamma}$.

From the above considerations, it follows that, in order to obtain the mass of the solution, only the $\mathcal{E}^{t}_{t}$ component of the electrical part of the Weyl tensor is needed
\begin{eqnarray}
    \mathcal{E}^{t}_{t}&=&\frac{1}{d-2}W^{t r}_{t r}=\frac{1}{d-2}F^{t r}_{t r}+\mathcal{O}\left(\frac{1}{r^{2d}}\right)\,,\notag\\
    &=&\frac{(d-1)m}{2r^{d}}+\mathcal{O}\left(\frac{1}{r^{2d}}\right)\,.
\end{eqnarray}
As a direct consequence, the evaluation of the charge \eqref{QQCG} can be readily performed as
\begin{eqnarray}
Q_{\mathrm{QCG}}[\partial_{t}]&=&\frac{2\ell _{\mathrm{eff}}\,\Xi _{d}}{\kappa }\int\limits_{\Sigma}d^{d-1}x\sqrt{\gamma}\,r^{d-1}\,f\,\mathcal{E}_{t}^{t}\,,\notag\\
    &=&\Xi _{d}\,\frac{(d-1)m}{\kappa }\,\mathrm{Vol}(\Gamma)\,, \label{mass}
\end{eqnarray}
where $\mathrm{Vol}(\Gamma)=\int_{\Gamma}d^{d-1}x\sqrt{\gamma}$ is the volume of the transversal section. In particular, the above expression matches the mass of the topological black hole in QCG obtained by linearized methods in \cite{Deser:2011xc,Petrov:2005qt,Peng:2023mvd}.


\section{Conclusions}
\label{conc}

In this work we have studied the dynamics of the most general Quadratic Curvature Gravity (QCG) in general  $d+1$ dimensions. The bare cosmological constant gives rise to an effective cosmological constant in AdS spacetime, which is a functions of the couplings of the  Ricci-squared,  the Ricci scalar-squared   and the Gauss-Bonnet  terms. This effective cosmological constant introduces a new length scale in the theory which changes the asymptotics  of the theory.  We  review the analysis of propagating modes of this theory and the  conditions for their existence. We then find the most general  asymptotic black hole solution in    ($d+1$)-dimensional QCG away from the critical point, which is a generalization of the near-boundary form of Schwarzschild-Tangherlini-AdS black hole. The corresponding expansion clearly breaks down in the critical points, $\Xi_d=0$.

To calculate the Conformal Mass in QCG, we first evaluate the Weyl tensor on-shell. It is known that, to define
physical quantities for asymptotically AdS black holes like mass, angular momentum, or free energy, we have to eliminate possible infrared divergences.
To avoid such possible large-distance infinities, we give a detailed discussion on the Kounterterm charges in QCG with AdS asymptotics. Then we give the explicit form of the Conformal Mass in odd and even dimensions as a consistent truncation of the Kounterterm charges.

In particular, we find that the result \eqref {QQCG} matches the one found by Pang
in \cite{Pang:2011cs} (as a correction of the one in \cite{Okuyama:2005fg}).\footnote{Conformal Mass as defined in \cite{Pang:2011cs} may be written in our conventions by the substitutions $16\pi G\rightarrow \kappa $, $\ell\rightarrow \ell _{\mathrm{eff}}$, $n\rightarrow d+1$, $\alpha \rightarrow
\kappa \gamma $, $\gamma \rightarrow \kappa \alpha $ and $\beta \rightarrow
\kappa \beta $.} It also coincides with the Conformal Mass in the case of Einstein-Gauss-Bonnet
AdS gravity obtained in \cite{Jatkar:2015ffa} for $\alpha =\beta =0$ and $\kappa =16\pi G$, and for Einstein-AdS gravity in \cite{Jatkar:2014npa} for $\alpha=\beta=\gamma=0$.

It would be interesting to extent this work to the case of matter coupled to QCG. In this context, in four dimensions, the found metric \eqref{metric function} is reduced to the Schwarzschild-AdS black hole. This is because, in four dimensions,
 the curvature terms appearing in the action \eqref{QCGaction} are such that the field equations always admit an Einstein space as a solution. However, if a scalar field is coupled to curvature terms
and backreacts to the  Schwarzschild-AdS metric, a hairy black hole is generated (it scalarized the background metric) \cite{Doneva:2017bvd,Doneva:2018rou}. In the case of QCG  in five dimensions,  the metric \eqref{metric function} is not a trivial function but it contains the information of the curvature terms.
In case a backreacting solution of the scalar field to the metric \eqref{metric function} exists in five dimensions, the scalarized solution will
have the information of high curvature terms plus the scalar hair.

\section*{Acknowledgments}

This work has been funded by the FONDECYT Regular Grants 1230492 and 1231779, and ANILLO Grant ANID/ACT210100. Y.P.C. is supported by the ANID National M.Sc.~Scholarship No.~22210771.

\appendix

\section{Useful identities}
\label{Conventions}

\subparagraph{Totally anti-symmetric Kronecker delta of rank $p$.}

Throughout the text, it is often used the totally anti-symmetric Kronecker delta of order $p$,
\begin{equation}
\delta^{\mu_{1} \mu_{2}\cdots\mu_{p}}_{\nu_{1} \nu_{2}\cdots\nu_{p}}= \det[\delta^{\mu_{1}}_{\nu_{1}}\cdots\delta^{\mu_{p}}_{\nu_{p}}]\,.
\end{equation}
When considering the contraction of $k$ indices, it follows the relation
\begin{equation}
\delta^{\mu_{1}\cdots\mu_{k}\cdots\mu_{p} }_{\nu_{1}\cdots\nu_{k}\cdots\nu_{p} }\delta^{\nu_{1}}_{\mu_{1}}\cdots\delta^{\nu_{k}}_{\mu_{k}}=\frac{\left(N-p+k\right)!}{\left(N-p\right)!}\, \delta^{\mu_{k+1}\cdots\mu_{p} }_{\nu_{k+1}\cdots\nu_{p}}\,,
\end{equation}
 where  $k\leq p$ and $N$ is the range of the indices.

\subparagraph{Useful integrals.}
The following integral identities are useful in some derivations
\begin{equation} \label{b-a}
b^{p}-a^{p}=p(b-a)\int\limits_{0}^{1}du\,\left[ \rule{0pt}{13pt}u(b-a)+a%
\right] ^{p-1},\qquad p\geq 1\,,
\end{equation}%
\begin{equation}\label{u^2-1}
\int\limits_{0}^{1}du\,\left( u^{2}-1\right) ^{n-2}=-\frac{2n-1}{2n-2}
\int\limits_{0}^{1}du\,\left( u^{2}-1\right) ^{n-1}\,.
\end{equation}
The above formula defines a recursion relation, such that
\begin{equation} \label{parametric}
\int\limits_{0}^{1}du\,\left( u^{2}-1\right) ^{n-1}=\frac{(-1)^{n-1}2^{2n-2}(n-1)!^{2}}{(2n-1)!}\,.
\end{equation}

\newpage
\section{Asymptotic form of Schwarzschild-Tangherlini-AdS black hole}
\label{ASTAdS}
\hspace{0.1cm}

\subparagraph{Exact relations.}Consider the topological black hole ansatz
\begin{equation}
ds^{2}=-f^{2}(r)dt^{2}+\frac{dr^{2}}{f^{2}(r)}+r^{2}\gamma _{mn}(\varphi )\,d\varphi ^{m}d\varphi ^{n}\,,\quad
\varphi ^{m}\in \Gamma_k^{d-1}\,,
\end{equation}
where $\Gamma_k^{d-1}$ is the transversal section, labelled by the topological parameter $k$.

The different components of the Christoffel connection are given by
\begin{equation}\label{Christoffel}
\begin{array}{llllll}
\Gamma _{tr}^{t} & =\dfrac{\left( f^{2}\right)' }{2f^{2}}\,, &
\Gamma _{tt}^{r} & =\dfrac{1}{2}\,f^{2}\left( f^{2}\right)'\,,\qquad  & \Gamma _{rr}^{r} & =-\dfrac{\left( f^{2}\right) '}{2f^{2}}\,,\medskip  \\
\Gamma _{mn}^{r} & =-rf^{2}\gamma _{mn}\,,\qquad  & \Gamma _{rm }^{n} & =\dfrac{1}{r}\,\delta _{m}^{n}\,, & \Gamma _{mn}^{k} & =\Gamma _{mn}^{k}(\gamma )\,.
\end{array}
\end{equation}
This allows the computation of the components of the Riemann tensor
\begin{eqnarray}
R_{tr}^{tr} &=&-\dfrac{1}{2}\,(f^{2})''\,, \hspace{1cm}
R_{kl}^{mn}=-\frac{f^{2}-k}{r^{2}}\,\delta_{kl}^{mn}\,,\notag\\
R_{tm }^{tn}&=&R_{rm }^{rn}=-\frac{(f^{2})'}{2r}\,\delta
_m^n\,,
\end{eqnarray}
the Ricci tensor
\begin{eqnarray}
R_{t}^{t} &=&R_{r}^{r}=-\dfrac{1}{2r^{2}}\left[ r^{2}(f^{2})''+(d-1) r(f^{2})'\right]  \,, \notag \\
R_{m}^{n} &=&-\dfrac{1}{r^{2}}\,\delta _{m}^{n}\left[ r(f^{2})'+(d-2) \left( f^{2}-k\right) \right]  \,,
\end{eqnarray}
and the Ricci scalar
\begin{equation}
    R =-\dfrac{1}{r^{2}}\,\left[ r^{2}\left( f^{2}\right)''+2(d-1) r(f^{2})'+(d-1)(d-2)
\left( f^{2}-k\right) \right]\,.
\end{equation}
\subparagraph{Asymptotic expressions.} The asymptotic form of the metric function is such that it matches the one of Schwarzschild-AdS black holes. Indeed, as dictated by eq.\eqref{falloff},
\begin{equation}
f^{2}=k+\frac{r^{2}}{\ell_{\mathrm{eff}}^{2}}-\frac{m}{r^{d-2}}+\frac{p}{r^{2d-2}}+\mathcal{O}\left( \frac{1}{r^{2d-1}}\right) \,.
\end{equation}
In order to evaluate the different components of the curvature tensors, it is useful to write the following expansions of the metric function and its derivatives
\begin{eqnarray}
\frac{\left( f^{2}-k\right) ^{2}}{r^{2}} &=&\frac{r^{2}}{\ell _{\mathrm{eff}}^{4}}-\frac{2m}{\ell_{\mathrm{eff}}^{2}r^{d-2}}+\left(m^2+\frac{2p}{\ell _{\mathrm{eff}}^{2}}\right) \frac{1}{r^{2d-2}}+\mathcal{O}\left(\frac{1}{r^{2d-1}}\right) \,, \notag \\
r(f^{2})' &=&\frac{2r^{2}}{\ell _{\mathrm{eff}}^{2}}+\frac{(d-2)m}{r^{d-2}}-\frac{2(d-1)p}{r^{2d-2}}+\mathcal{O}\left( \frac{1}{r^{2d-1}}\right) \,,  \\
r^{2}(f^{2})'' &=&\frac{2r^{2}}{\ell _{\mathrm{eff}}^{2}}-\frac{(d-1)(d-2)m}{r^{d-2}}+\frac{2(d-1)(2d-1)p}{r^{2d-2}}+\mathcal{O}\left(\frac{1}{r^{2d-1}}\right) \,. \notag
\end{eqnarray}
It is then straightforward to find the asymptotic form of the Riemann curvature tensor
\begin{eqnarray} \label{Riemann onshell}
R_{tr}^{tr} &=&-\frac{1}{\ell _{\mathrm{eff}}^{2}}+\frac{(d-1)(d-2)m}{2r^{d}}-\frac{(d-1)(2d-1)p}{r^{2d}}+\mathcal{O}\left( \frac{1}{r^{2d+1}}\right) ,  \notag  \\
R_{tm }^{tn} &=&R_{rm }^{rn}=\left[ -\frac{1}{\ell_{\mathrm{eff}}^{2}}-\frac{(d-2)m}{2r^{d}}+\frac{(d-1)p}{r^{2d}}+\mathcal{O}\left( \frac{1}{r^{2d+1}}\right) \right]
\delta _m^{n}\,,\notag\\
R_{kl}^{mn}&=&\left[ -\frac{1}{\ell _{\mathrm{eff}}^{2}}+\frac{m}{r^{d}}-\frac{p}{r^{2d}}+\mathcal{O}\left( \frac{1}{r^{2d+1}}\right) \right]
\,\delta_{kl}^{mn}\,,
\end{eqnarray}
as well as for the Ricci tensor
\begin{eqnarray}
R_{t}^{t} &=&R_{r}^{r}=-\frac{d}{\ell _{\mathrm{eff}}^{2}}-\frac{d(d-1)p}{r^{2d}}+\mathcal{O}\left( \frac{1}{r^{2d+1}}\right) \,, \notag \\
R_{m}^{n} &=&\delta _{m}^{n}\left[ - \frac{d}{\ell_{\mathrm{eff}}^{2}} +\frac{dp}{r^{2d}}+\mathcal{O}\left( \frac{1}{r^{2d+1}}\right) \right] \,,
\end{eqnarray}
and the Ricci scalar
\begin{eqnarray}
R &=&-\frac{d(d+1) }{\ell _{\mathrm{eff}}^{2}}-\frac{d(d-1)p}{r^{2d}}+\mathcal{O}\left( \frac{1}{r^{2d+1}}\right) \,.
\end{eqnarray}
On the other hand, in the component of the tensor $P^{t}_{t}$ in eq.\eqref{Psi} there are contributions of quadratic order in the curvature and of fourth order in derivatives. In particular, from eq.\eqref{G0}, one may render explicit the expression
\begin{eqnarray}
\frac{2r^{d-1}}{d-1}\,2RG_{t}^{t}|_{\Lambda _{0}=0} &=&2R\left[ \rule{0pt}{14pt}r^{d-2}\left( f^{2}-k\right) \right]' \\
&=&\left[ -\frac{2d(d+1)}{\ell _{\mathrm{eff}}^{4}}\,r^{d}-\frac{4dp}{\ell _{\mathrm{eff}}^{2}r^{d}}+\mathcal{O}\left( \frac{1}{r^{d+1}}\right) \right]'.  \notag
\end{eqnarray}
In turn, the quadratic term of the Ricci scalar accepts the expansion
\begin{equation}
\frac{r^{d-1}}{d-1}\,R^{2}=\frac{d(d+1)}{(d-1)\ell _{\mathrm{eff}}^{2}}\,\left[ \frac{d+1}{\ell _{\mathrm{eff}}^{2}}\,r^{d}-\frac{2(d-1)p}{r^{d}}+\mathcal{O}\left( \frac{1}{r^{d+1}}\right) \right]'.
\end{equation}
The corresponding contraction between the Riemann and the Ricci tensors are
\begin{equation}
\frac{4r^{d-1}}{d-1}\,R_{t\lambda }^{t\sigma }R_{\sigma }^{\lambda } =\frac{4r^{d-1}}{d-1}\,\left( R_{tr}^{tr}R_{r}^{r}+R_{tn}^{tm}R_{m}^{n}\right) =\frac{4d}{\ell _{\mathrm{eff}}^{2}}\,\left[ \frac{r^{d}}{(d-1)\ell _{\mathrm{eff}}^{2}}-\frac{p}{r^{d}}+\mathcal{O}\left( \frac{1}{r^{d+1}}\right) \right]',
\end{equation}
and the Ricci-squared term
\begin{equation}
-\frac{r^{d-1}}{d-1}\,R_{\lambda }^{\sigma }R_{\sigma }^{\lambda }=-\frac{r^{d-1}}{d-1}\,\left[ 2(R_{t}^{t})^{2}+R_{n}^{m}R_{m}^{n}\right] =-\frac{d}{\ell _{\mathrm{eff}}^{2}}\,\left[ \frac{d+1}{(d-1)\ell _{\mathrm{eff}}^{2}} \,r^{d}-\frac{2p}{r^{d}}+\mathcal{O}\left( \frac{1}{r^{d+1}}\right) \right]'.
\end{equation}

The derivative terms contribute in the following way,
\begin{equation}
\frac{2r^{d-1}}{d-1}\,\square R
=\left( \frac{2r^{d-1}}{d-1}\,f^{2}R'\right)'=\left[ \frac{4d^{2}p}{\ell _{\mathrm{eff}}^{2}r^{d}}+\mathcal{O}\left( \frac{1}{r^{d+1}}\right) \right]',
\end{equation}
and similarly for the Ricci tensor
\begin{equation}
\frac{2r^{d-1}}{d-1}\,\square R_{t}^{t}=\frac{2r^{d-1}}{d-1}\,\left[
f^{2}\left( R_{t}^{t}\right)''+\dfrac{1}{2}\,\left(
f^{2}\right)'\,\left( R_{t}^{t}\right)'\right] =\left[
\frac{8d^{2}p}{\ell _{\mathrm{eff}}^{2}r^{d}}+\mathcal{O}\left( \frac{1}{r^{d+1}}\right) \right]'\,.
\end{equation}
Finally, the Einstein's tensor takes the form
\begin{equation}
    \frac{2r^{d-1}}{d-1}G^{t}_{t}=\left[\frac{\ell_{0}-\ell_{\mathrm{eff}}^{2}}{\ell_{\mathrm{0}}^{2}\ell_{\mathrm{eff}}^{2}}r^{d}-m+\frac{p}{r^{2}} +\mathcal{O}\left(\frac{1}{r^{d+1}}\right)\right]'\,,
\end{equation}
whereas the Lanczos tensor reads
\begin{equation}
    \frac{2r^{d-1}}{d-1}H^{t}_{t}=-(d-2)(d-3)\left[\frac{r^{d}}{\ell^{4}_{\mathrm{eff}}}-\frac{2m}{\ell^{2}_{\mathrm{eff}}}+ \left(m^{2}+\frac{2p}{\ell^{2}_{\mathrm{eff}}} \right)\frac{1}{r^{d}}+\mathcal{O}\left(\frac{1}{r^{d+1}}\right)\right]'\,.
\end{equation}


\bibliographystyle{unsrt}

\end{document}